\documentclass[%
 twocolumn, 
superscriptaddress,
 amsmath,amssymb,
 aps, physrev,
]{revtex4-2}
\usepackage[T1]{fontenc}

\usepackage{longtable}
\usepackage{graphicx}
\usepackage{dcolumn}
\usepackage{bm}
\usepackage{enumitem} 
\usepackage{hyperref}
\usepackage{xcolor}

\begin{document}

\title{\textbf{Improved $^{94}$Mo neutron resonance parameters from neutron capture and transmission measurements at n\_TOF and GELINA} 
}

\author{%
R.~Mucciola$^{1,2,*}$, %
S.~Cristallo$^{1,3}$, %
S.~Kopecky$^{4}$, %
N.~Liu$^{5}$, %
C.~Massimi$^{6,7}$, %
A.~Mengoni$^{8,6}$, %
A.~Manna$^{6,7}$, %
C.~Paradela$^{4}$, %
P.~Schillebeeckx$^{4}$, %
G.~Sibbens$^{4}$
% ...
}
\collaboration{The n\_TOF Collaboration (www.cern.ch/ntof)}
\noaffiliation

\affiliation{{\small%
$^{}$Istituto Nazionale di Fisica Nucleare, Sezione di Perugia, Italy\\
$^{2}$Dipartimento di Fisica e Geologia, Universit\`{a} di Perugia, Italy\\
$^{3}$Istituto Nazionale di Astrofisica - Osservatorio Astronomico d'Abruzzo, Italy\\
$^{4}$European Commission, Joint Research Centre (JRC), Geel, Belgium\\
$^{5}$Insitute for Astrophysical Research, Boston University, Boston, USA\\
$^{6}$Istituto Nazionale di Fisica Nucleare, Sezione di Bologna, Italy\\
$^{7}$Dipartimento di Fisica e Astronomia, Universit\`{a} di Bologna, Italy\\
$^{8}$Agenzia nazionale per le nuove tecnologie, l'energia e lo sviluppo economico sostenibile (ENEA), Italy\\
$^{9}$European Organization for Nuclear Research (CERN), Switzerland\\
$^{10}$Centro de Investigaciones Energ\'{e}ticas Medioambientales y Tecnol\'{o}gicas (CIEMAT), Spain\\
$^{11}$Istituto Nazionale di Fisica Nucleare, Sezione di Pavia, Italy\\
$^{12}$Department of Physics, University of Pavia, Italy\\
$^{13}$INFN Laboratori Nazionali del Sud, Catania, Italy\\
$^{14}$University of Lodz, Poland\\
$^{15}$Instituto de F\'{\i}sica Corpuscular, CSIC - Universidad de Valencia, Spain\\
$^{16}$University of Manchester, United Kingdom\\
$^{17}$CEA Irfu, Universit\'{e} Paris-Saclay, F-91191 Gif-sur-Yvette, France\\
$^{18}$Helmholtz-Zentrum Dresden-Rossendorf, Germany\\
$^{19}$Horia Hulubei National Institute of Physics and Nuclear Engineering, Romania\\
$^{20}$Department of Physics, Faculty of Science, University of Zagreb, Zagreb, Croatia\\
$^{21}$University of Santiago de Compostela, Spain\\
$^{22}$Universitat Polit\`{e}cnica de Catalunya, Spain\\
$^{23}$Istituto Nazionale di Fisica Nucleare, Sezione di Trieste, Italy\\
$^{24}$Department of Physics, University of Trieste, Italy\\
$^{25}$National Technical University of Athens, Greece\\
$^{26}$Istituto Nazionale di Fisica Nucleare, Sezione di Torino, Italy\\
$^{27}$Department of Physics, University of Torino, Italy\\
$^{28}$Istituto Nazionale di Fisica Nucleare, Sezione di Bari, Italy\\
$^{29}$Universidad de Sevilla, Spain\\
$^{30}$Goethe University Frankfurt, Germany\\
$^{31}$Physikalisch-Technische Bundesanstalt (PTB), Bundesallee 100, 38116 Braunschweig, Germany\\
$^{32}$Instituto Superior T\'{e}cnico, Lisbon, Portugal\\
$^{33}$Paul Scherrer Institut (PSI), Villigen, Switzerland\\
$^{34}$University of Ioannina, Greece\\
$^{35}$Affiliated with an institute covered by a cooperation agreement with CERN\\ 
$^{36}$University of Granada, Spain\\
$^{37}$Istituto Nazionale di Fisica Nucleare, Sezione di Roma1, Roma, Italy\\
$^{38}$University of York, United Kingdom\\
$^{39}$TU Wien, Atominstitut, Stadionallee 2, 1020 Wien, Austria\\
$^{40}$Japan Atomic Energy Agency (JAEA), Tokai-Mura, Japan\\
$^{41}$Charles University, Prague, Czech Republic\\
$^{42}$School of Physics and Astronomy, University of Edinburgh, United Kingdom\\
$^{43}$INFN Laboratori Nazionali di Legnaro, Italy\\
$^{44}$Dipartimento Interateneo di Fisica, Universit\`{a} degli Studi di Bari, Italy\\
$^{45}$Consiglio Nazionale delle Ricerche, Bari, Italy\\
$^{46}$INFN Laboratori Nazionali di Frascati, Italy\\
$^{47}$Istituto Nazionale di Fisica Nucleare, Sezione di Catania, Italy\\
$^{48}$Department of Physics and Astronomy, University of Catania, Italy\\
$^{49}$Department of Physics and Astronomy, Uppsala University, Box 516, 75120 Uppsala, Sweden\\
\texorpdfstring{$^{*}$}{*} present affiliation: INFN-Bari\\
}}

\date{\today}

\begin{abstract}
We report high-resolution measurements of the $^{94}\mathrm{Mo}(\mathrm{n},\gamma)^{95}\mathrm{Mo}$ cross section in the neutron energy range from a few eV up to about 250~keV, performed at the n\_TOF facility (CERN), and of the $^{94}\mathrm{Mo}(\mathrm{n},\mathrm{tot})$ cross section up to 32~keV, measured at GELINA (JRC Geel). A combined R-matrix analysis of capture and transmission data yields significantly improved neutron-resonance parameters for $^{94}$Mo. A total of 186 resonances were observed in this analysis of which 127 reported here for the first time. The resulting Maxwellian-averaged cross section at stellar temperatures relevant to the slow neutron-capture process  (s-process) is approximately 25\% lower than previous evaluations and literature values. To assess the astrophysical impact of the revised cross section, we performed s-process nucleosynthesis calculations for low-mass asymptotic giant branch stars. Despite the substantial reduction in the $^{94}\mathrm{Mo}(\mathrm{n},\gamma)^{95}\mathrm{Mo}$ rate, the final yields vary by only 3-4\%. This demonstrates that the isotopic budget of $^{94}$Mo in AGB stars is primarily governed by the branching flow at $^{94}\mathrm{Nb}$ rather than by the neutron-capture destruction on $^{94}$Mo itself. The new cross
section thus helps disentangle nuclear cross-section uncertainties from branching-flow effects in the s-process production of $^{94}$Mo.
\end{abstract}

\maketitle

\section{\label{sec:Introduction}Introduction}
The molybdenum isotopic chain provides a stringent test for nucleosynthesis models because its
seven stable isotopes originate from different astrophysical processes~\cite{Arlandini99}. The neutron-deficient isotopes $^{92}$Mo and $^{94}$Mo (mostly) are commonly associated with the p-process (i.e., the photodisintegration or $\gamma$-process), while $^{96}$Mo is produced exclusively by the slow neutron-capture process (s-process), $^{100}$Mo is
predominantly of rapid neutron-capture process (r-process) origin, and the remaining isotopes ($^{95}$Mo,
$^{97}$Mo and $^{98}$Mo) receive mixed s- and r-process contributions. This diversity makes Mo a sensitive probe
of the interplay between neutron-capture and charged-particle processes in stellar environments. Among
these isotopes, $^{94}$Mo is of special interest because its abundance can be affected by the s-process
contribution coming from the branching at $^{94}$Nb~\cite{Lugaro2003}, making precise knowledge of neutron-capture cross sections essential for reliable nucleosynthesis modeling.

Isotopic compositions measured in presolar silicon carbide (SiC) grains provide direct constraints on
stellar nucleosynthesis~\cite{LIU2025113}. These microscopic stardust particles, isolated from primitive meteorites, retain the isotopic signatures of their parent stars. The majority of presolar SiC grains are inferred to originate from low-mass asymptotic giant branch (AGB) stars~\cite{Liu2022}, the main stellar site of the s-process~\cite{Gallino_1998}. Comparisons between grain isotopic data and AGB model predictions therefore depend sensitively on the adopted nuclear reaction network, in particular on reliable neutron-capture cross sections. Reported discrepancies between predicted and measured Mo isotopic abundances~\cite{Liu19,Stephan2019ApJ...877..101S,Palmerini2021ApJ...921....7P} have underscored the need
for improved nuclear data, including an accurate determination of the $^{94}$Mo(n,$\gamma$)$^{95}$Mo cross section.

Along the s-process path, $^{93}$Nb captures a neutron to form radioactive $^{94}$Nb, which can either $\beta$-decay to $^{94}$Mo or capture an additional neutron. The final $^{94}$Mo abundance therefore depends on
the competition between neutron capture and $\beta$-decay at $^{94}$Nb. Quantifying this branching requires precise neutron-capture cross sections of both $^{94}$Mo and $^{94}$Nb in order to disentangle nuclear-rate effects from stellar-model uncertainties~\cite{Cescutti18}. Its astrophysical relevance is further supported by spectroscopic observations of Nb in intrinsic AGB stars and in barium stars enriched through mass transfer from a former AGB companion~\cite{Shetye20,Busso95}.

The existing experimental database for the $^{94}$Mo(n,$\gamma$) reaction is limited in precision and exhibits significant inconsistencies among evaluated libraries. A recent study of previously published measurements~\cite{Mucciola22} revisited the neutron-resonance parameters of Mo isotopes below $E_n= 5$~keV and identified discrepancies and impact of possible systematic effects in earlier experiments~\cite{Weigmann67,Musgrove76,Wang08,Wynchank68,Leinweber10}. In particular, that study~\cite{Mucciola22} demonstrated that current evaluated nuclear data files fail to adequately reproduce the available $\text{n}+^{94}$Mo data. Consistent with this picture, the recommended Maxwellian-averaged cross section (MACS) derived from different evaluations shows noticeable differences. According to the KADoNiS v0.3 compilation~\cite{Dillmann2009_KADoNiS_v0.3}, the dominant source of MACS used in astrophysical calculations, the recommended MACS for $^{94}$Mo(n,$\gamma$) at $kT = 30$ keV is $102 \pm 20$ mb~\cite{Bao2000}, while at temperatures relevant to low-mass AGB stars ($kT = 8$ keV), the MACS values derived from different evaluated libraries (210 mb for ENDF/B-VIII.1~\cite{ENDF8.1}, 204 mb for JEFF-4.0~\cite{JEFF4.0}, and 230 mb for JENDL-5.0~\cite{JENDL5.0}) differ at the 10–20\% level. Such discrepancies translate directly into uncertainties in s-process abundance predictions and complicate the interpretation of branching effects at $^{94}$Nb. 

\begin{figure}[t]
    \centering
    \includegraphics[width=\columnwidth]{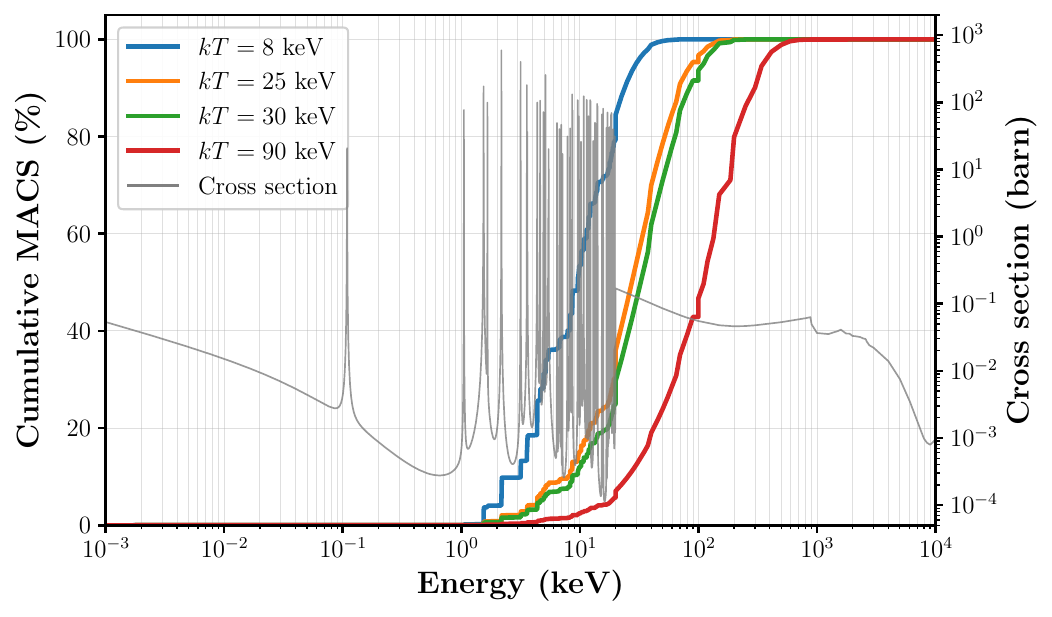}
    \caption{(Color online) The cross section of $^{94}$Mo(n,g) reaction as available in ENDF/B-VIII.1 and its cumulative contribution to the MACS at different temperatures.}\label{MACS_ENDF}
\end{figure}

The energy region of interest for the $^{94}$Mo(n,$\gamma$) cross-section in astrophysics calculations can be estimated from the cumulative of the MACS as a function of neutron energy for different values of kT.
Fig.~\ref{MACS_ENDF} illustrates that neutron energies between approximately 1 and 20 keV dominate the MACS at s-process temperatures and 95\% of the MACS at 90~keV can be obtained from cross-sections below 250~keV. The spread among current evaluations in this astrophysically relevant energy window, together with the deficiencies identified in previous analyses~\cite{Mucciola22}, clearly demonstrates the need for new high-resolution capture and transmission measurements. The present work addresses this requirement using a highly enriched $^{94}$Mo sample and a combined
experimental approach.

\section{\label{sec:Samples}Sample details}
To reduce the contamination from other molybdenum isotopes, a highly enriched Mo sample was used. The molybdenum sample was enriched at 98.97\% in $^{94}$Mo, all other molybdenum isotopes were present at levels below 0.7\%. The isotopic abundances of the different Mo isotopes in the sample are reported in Table~\ref{MoEnrichment}. 

\begin{table}[h] 
\caption{Declared abundances of the enriched molybdenum powder.}
\begin{tabular} {ccccccc} \hline \label{MoEnrichment}

$^{92}$Mo   & $^{94}$Mo    & $^{95}$Mo    & $^{96}$Mo    & $^{97}$Mo   & $^{98}$Mo   & $^{100}$Mo  \\ \hline
0.63\% & 98.97\% & 0.36\%  & 0.01\%  & 0.01\% & 0.01\% & 0.01\% \\
\hline
\end{tabular}
\end{table}

The sample consisted of metallic Mo powder, pressed into a self-sustaining disk with 20 mm diameter and sealed under vacuum in thin plastic bags, to avoid loss of material. The mass of the sample was $1.9526(3)$~g, corresponding to an areal density of $3.960(4)\times 10^{-3}$~at/b. While this large thickness increases the probability of multiple neutron interactions in the sample and therefore requires sizable multiple scattering corrections, it ensures an adequate signal-to-background ratio. To check the enrichment of the sample and the effect of the grain size in the experimental measurements, two additional $^{\text{nat}}$Mo samples were prepared using metallic powder with $350~\mu\text{m}$ and $<5~\mu\text{m}$ grain size, respectively. The main characteristics of all the molybdenum samples are reported in Table~\ref{MoSamples}. In addition to the molybdenum sample, a gold ($5.834 \times 10^{-4}$~at/b), lead ($3.338 \times 10^{-3}$~at/b), and carbon ($8.703\times10^{-3}$ at/b) sample were used for normalization and background determination. All these samples have the same shape and intercepted the same fraction of the neutron beam. 

\begin{table}[h] 
\caption{Mass and areal densities of the $^{94}$Mo enriched and $^{\text{nat}}$Mo samples.}
\begin{tabular}{ccc} \hline \label{MoSamples}
Sample & Mass & Areal density\\ 
ID &  (g) & (at/b)\\ \hline
$^{94}$Mo  &$1.9526(3)$&  $3.960(4)\times10^{-3}$  \\
$^{\text{nat}}$Mo--$5 \mu\text{m}$ & $2.014(2)$ & $4.006(4)\times10^{-3}$\\
$^{\text{nat}}$Mo--$350 \mu\text{m}$ & $1.989(2)$ & $3.958(4)\times10^{-3}$\\
\hline
\end{tabular}
\end{table}

\section{\label{sec:Capture}Capture measurement}
The radiative capture cross-section measurement was performed in 2022 at the n\_TOF 185 m experimental area (EAR1), exploiting the time-of-flight (TOF) technique and the total energy detection method~\cite{Schillebeeckx12}. The measurement covered neutron energies from a few eV up to $E_n = 250$~keV. The cross section was determined through an R-matrix analysis up to 76~keV. At higher energies the resolution of the facility does not allow the use of this approach. For energies between 50~keV and 250~keV the capture cross section was obtained using the unresolved resonance region (URR) approach, using the same procedure described in previous work \cite{Sosnin23}. The cross section in the energy range of 50~keV to 76~keV was obtained using both approaches, and the results can be compared to check for the consistency of the two techniques. 

\subsection{Experiment at n\_TOF - EAR1}\label{sec2.1}
The n\_TOF facility~\cite{Guerrero13} features two broad-spectrum neutron beam lines for time-of-flight studies. Neutrons are generated by 20~GeV/c protons from the CERN PS impinging on a massive lead-target assembly~\cite{Esposito21} serving as spallation target. A 4~cm layer of borated water (H$_2$O with 1.28\% H$_3$BO$_3$) moderates the initially fast spectrum, yielding neutron energies from a few meV to the GeV region. Pulses are delivered with a frequency lower than 1~Hz, preventing overlap between neutrons of consecutive bunches.

The capture measurement reported here was performed in EAR1, the farthest experimental area, because EAR1 provides the best energy resolution for the study of resonance cross sections. In fact, the energy resolution is $\Delta E_n / E_n=4.3\times 10^{-4}$ at $E_n=100$~eV and $\Delta E_n / E_n=1.1\times10^{-3}$ at $E_n=10$ keV~\cite{Guerrero13,Esposito21}. 

During the experiment, we used a capture setup to measure the $\gamma$ rays originating from the $^{94}$Mo(n,$\gamma$) reaction, and we monitored the neutrons impinging on the sample with a flux monitor.

The capture setup consisted of 4 liquid scintillation detectors based on deuterated benzene (C$_6$D$_6$), each with a volume of approximately 1~L. These detectors, which are well known for their suitability in (n,$\gamma$) measurements~\cite{Schillebeeckx12}, were specially refined in previous work~\cite{Mastinu13} to suppress background originating from sample-scattered neutrons reaching the detector. The low neutron sensitivity achieved with these detectors is particularly important for the present measurement, as the ratio of the radiative capture to elastic cross section can be as small as $\sigma_\gamma/\sigma_{el}\approx 10^{-3}$.

The detectors were positioned opposite each other, forming a ${135}^\circ$ angle with respect to the neutron beam axis, at a distance of roughly 8~cm from the center of the Mo sample. This geometrical configuration minimized the background due to scattered in-beam $\gamma$ rays and reduced systematic effects due to the anisotropy in the angular distribution from primary $\gamma$-rays following p-wave neutron resonances.

The number of neutrons impinging on the sample was monitored by means of a low-mass, $^{6}$Li-based, neutron detector~\cite{Marrone04}, while the energy dependence of the neutron flux was derived in a dedicated study~\cite{FLUX2025}. More specifically, the evaluated flux is obtained through a combination of dedicated measurements performed with several detector systems based on neutron cross-section standards defined in ~\cite{Carlson09}. 

For both capture and flux systems, signal digitization was carried out using TELEDYNE SP-Devices flash-ADC modules with 14-bit resolution. Each channel features 512~MB of on-board memory, allowing continuous recording over time windows of up to 100~ms, thereby covering neutron energies down to about 18~meV in EAR1. Four ADC channels were dedicated to the C$_6$D$_6$ detectors operating at a sampling rate of 1~GSample/s, while four additional channels were assigned to the neutron flux monitors, sampled at 62.5~MSample/s.

Neutron kinetic energy, $E_n$, cannot be simply obtained as the ratio of the geometrical flight path $L_0$ to the measured time of flight $\mathrm{TOF}_m$, because $\mathrm{TOF}_m$ also includes the moderation time inside the target assembly. As discussed in Ref.~\cite{Guerrero13}, the moderation-time distribution is usually expressed as an equivalent distance $\lambda(E_n)$, which depends on the neutron energy. The effective flight path is therefore $L(E_n) = L_0 + \langle \lambda(E_n)\rangle$
where $\langle \lambda(E_n) \rangle$ is the average of the distribution obtained from Monte Carlo simulations. This average, about 19~cm, varies only weakly with neutron energy in the range of interest. The kinetic energy can be then calculated as
\begin{equation}
E_n = \frac{1}{2} m_n \left( \frac{L_0+\langle \lambda(E_n)\rangle}{\mathrm{TOF}_m}\right)^2
\label{eq:kinetic_energy}
\end{equation}
where $m_n$ represent the neutron mass, $\mathrm{TOF}_m$ is the measured time-of-flight defined as $\mathrm{TOF}_m=t-(t_\gamma - L_0/c+\tau_0)$ where $t$ is the signal time, $t_\gamma$ is the time of arrival of the gamma-flash and $\tau_0$ is a time offset.
The kinetic energy is then evaluated through a recursive procedure that converges after a few iterations. The value $L_0 = 183.92(4)\,\text{m}$ was determined by fitting the well-known low-energy resonances of $^{197}$Au from the JEFF-4.0 evaluation~\cite{JEFF4.0}. Moreover, a time offset of $\tau_0 = -26.22(3)\,\text{ns}$ was obtained by fitting the $^{197}$Au resonances up to 1.5~keV.

\subsection{Capture-Yield Determination}
The neutron-capture yield $Y(E_n)$, defined as the fraction of the neutron beam inducing a capture reaction in the sample, was determined following Ref.~\cite{Schillebeeckx12}:
\begin{equation}
Y(E_n) = N\frac{1}{S_n + E_n\frac{ A}{A+1}} \frac{ C_w(E_n) - B_w(E_n) }{
\phi_n(E_n)}.
\label{eq:yield}
\end{equation}

Here, $C_w(E_n)$ is the weighted C$_6$D$_6$ counting rate for the sample under investigation, i.e. either $^{94}$Mo or $^{197}$Au, $N$ is an energy-independent normalization constant, and $S_n$ denotes the neutron separation energy of the compound nucleus ($S_n = 6.15$ MeV and $S_n = 7.37$ MeV for $^{198}$Au and $^{95}$Mo, respectively). The term $S_n + E_n\frac{ A}{A+1}$ represent the detector efficiency obtained when using the TED technique~\cite{Schillebeeckx12}. The term $A$ represents the mass number of the target nucleus, $B_w(E_n)$ are the weighted background counts, and $\phi_n(E_n)$ is the neutron fluence.

To implement the total-energy detection method, the C$_6$D$_6$ array was used together with the Pulse Height Weighting Technique (PHWT)~\cite{Abbondanno2004,Borella07}. This technique ensures that the detection efficiency for a capture event is directly proportional to the total $\gamma$-ray energy released in the event. The weighting functions (WFs) were calculated on the basis of Geant4~\cite{Agostinelli03} Monte Carlo simulations. More specifically, the response of the detection setup was first simulated for mono-energetic $\gamma$ rays, and the resulting spectra were then convoluted with the detector resolution. The WFs were finally obtained by considering only events above the detection threshold, set to 180 keV for the first three detectors, and 240 keV for the last one. Since both the discrimination threshold and the detector resolution significantly affect the quality of the WFs, particular care was devoted to determining the experimental resolution as well as the energy calibration of the detectors.

The iterative procedure used for this purpose consisted of: i) extracting the energy resolution from experimental spectra acquired with standard $^{137}$Cs and $^{88}$Y sources, as well as a composite Am--Be $\gamma$-ray source; ii) broadening the simulated spectra using this energy-resolution function until reproducing the measured spectra; iii) performing a fine calibration of the detectors, obtained from the best agreement between simulated and measured spectra.

After applying the WFs to the experimental C$_6$D$_6$ counting rates---thus obtaining $C_w(E_n)$---the normalization $N$ in Eq.~\ref{eq:yield} was determined using the saturated-resonance technique~\cite{Schillebeeckx12} applied to the 4.9~eV s-wave resonance in $^{197}$Au(n,$\gamma$).

The determination of the $^{94}$Mo(n,$\gamma$) cross section starting from the capture yield is described in~\ref{sec:RSA} for the energy region up to 76~keV and in~\ref{sec:AVERAGE} for energies between 50 and 250~keV. In the former, the cross section is parameterized in terms of resonance parameters, while in the latter it is compared with a statistical model calculation based on average resonance parameters.  

\subsection{Background subtraction} \label{Background subtraction}
The background in the capture data, $B_w$ in Eq.~\ref{eq:yield}, was studied through dedicated measurements to evaluate its main components: i) time-independent background, ii) neutron beam interactions with anything other than the sample, iii) sample-scattered neutrons, and iv) in-beam $\gamma$ rays. 

The first component, due to ambient radioactivity and activation of experimental materials, was determined from a beam-off measurement. Activation of the Mo samples is negligible, as the capture product is the stable $^{95}$Mo isotope. The second component was determined with an empty-sample-holder measurement, accounting for beam-related effects unrelated to the sample. The third, due to $\gamma$ rays from sample-scattered neutrons captured in surrounding materials, was evaluated using a graphite sample; the empty-sample counts, normalized to neutron intensity, were subtracted from the spectra. The fourth component, mainly 2.2 MeV and 0.48 MeV $\gamma$ rays from neutron capture in H and B of the moderator, was estimated with the lead measurement. The background component given by the in-beam $\gamma$ rays was first disentangled from the counts coming from neutron scattering in the lead sample by subtracting the spectra from graphite, normalized to neutron intensity. This background contributes above 300~eV (TOF$_m \sim 7.7 \times 10^5$~ns) and reflects the combined neutron slowing-down and flight-path effects. 

\begin{figure}[h]
    \centering
    \includegraphics[width=\columnwidth]{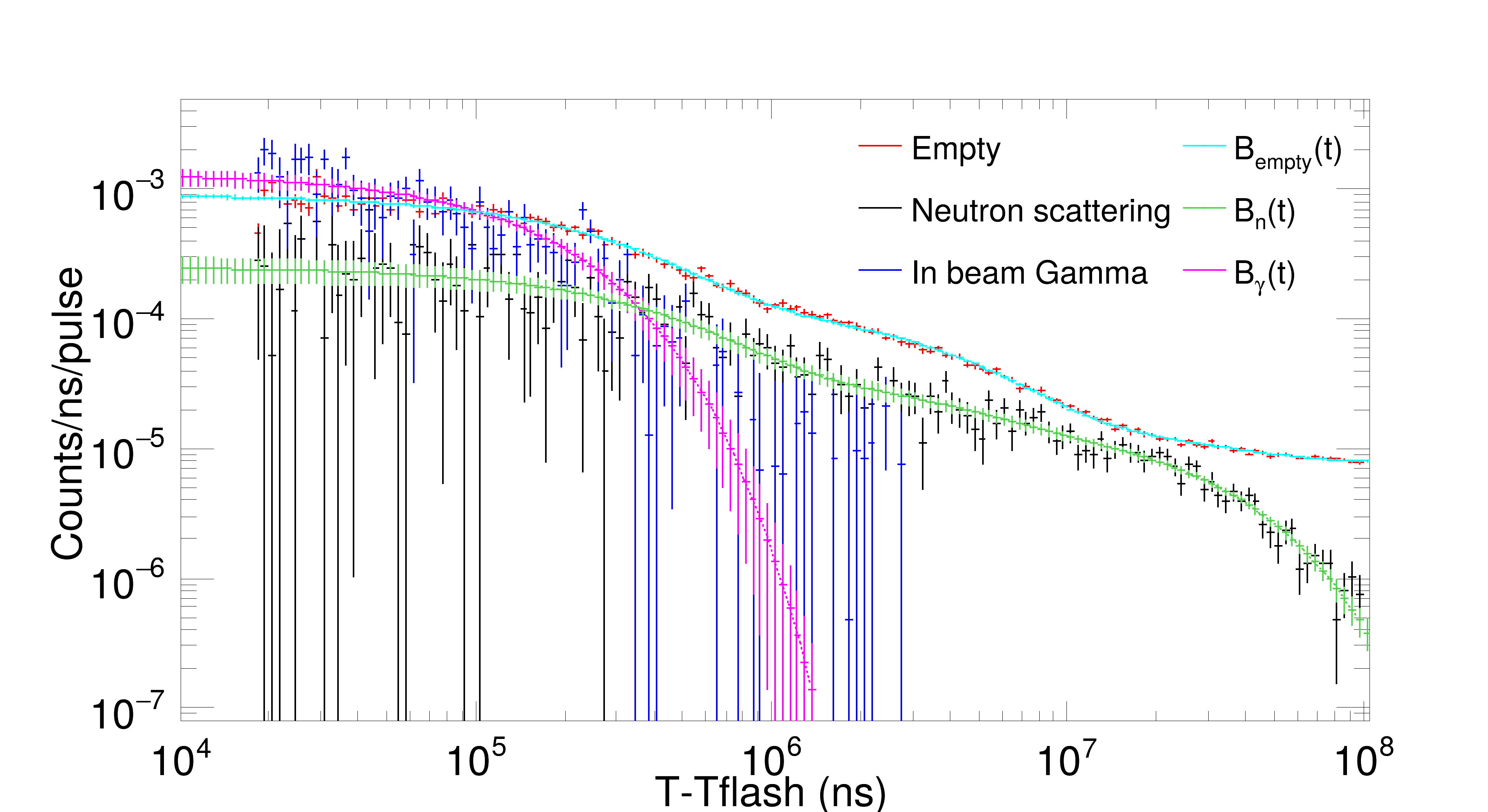}
    \caption{Experimental spectra for empty background (red), neutron scattering (black), and in-beam $\gamma$-rays (blue) together with the fitted curves. The fitted background components are shown together with their confidence intervals. The region of interest is between $2.5\times10^4$ ns and $3.0\times10^6$ ns, corresponding to neutron kinetic energy of 270 keV and 20 eV, respectively.}\label{fig:BKG_fit}
\end{figure}

In order to remove the statistical fluctuations in the shapes of the time-dependent background components, we performed a fit of the experimental TOF spectra. The different components were fitted using combinations of exponential curves to better reproduce the experimental histograms. The uncertainty on the background components associated with the fit was evaluated using the confidence interval of the fit. To obtain the total background for $^{94}$Mo, the neutron scattering and in-beam $\gamma$-rays background level were adjusted using dedicated measurements performed with a $^{209}$Bi black resonance filter in the beam and adjusting the background to the black resonances at 800 eV, 2 keV, and 12 keV. The total background obtained with the black resonance technique~\cite{Schillebeeckx12} can be expressed as:

\begin{equation}
B_w(t)=B_0+B_{empty}(t)+ f_n R_n B_n (t)+f_\gamma R_\gamma B_\gamma(t) .
\label{eq:background_filters}
\end{equation}
where individual terms correspond to the background sources (i)-(iv) mentioned at the beginning of the section, $f_\gamma$ and $f_n$ are the scaling factors for the in-beam $\gamma$-rays and the neutron scattering background components obtained using the black resonances, while $R_\gamma$ and $R_n$ are scaling factors defined as
\begin{equation}
R_\gamma=\frac{Z_{Mo}}{Z_{Pb}}\frac{n_{Mo}}{n_{Pb}}; R_n=\frac{\langle\sigma_{el,Mo}\rangle}{\langle\sigma_{el,C}\rangle}\frac{n_{Mo}}{n_C}
\label{eq:background_filters}
\end{equation}
The values of the scaling factors obtained from the fit of the black resonance dips are $f_\gamma=0.69(5)$ and $f_n=0.75(5)$, respectively. The experimental spectra for the three time-dependent background components together with the respective fit and confidence intervals are shown in Fig.\ref{fig:BKG_fit}

\begin{figure}[h]
    \centering
    \includegraphics[width=\columnwidth]{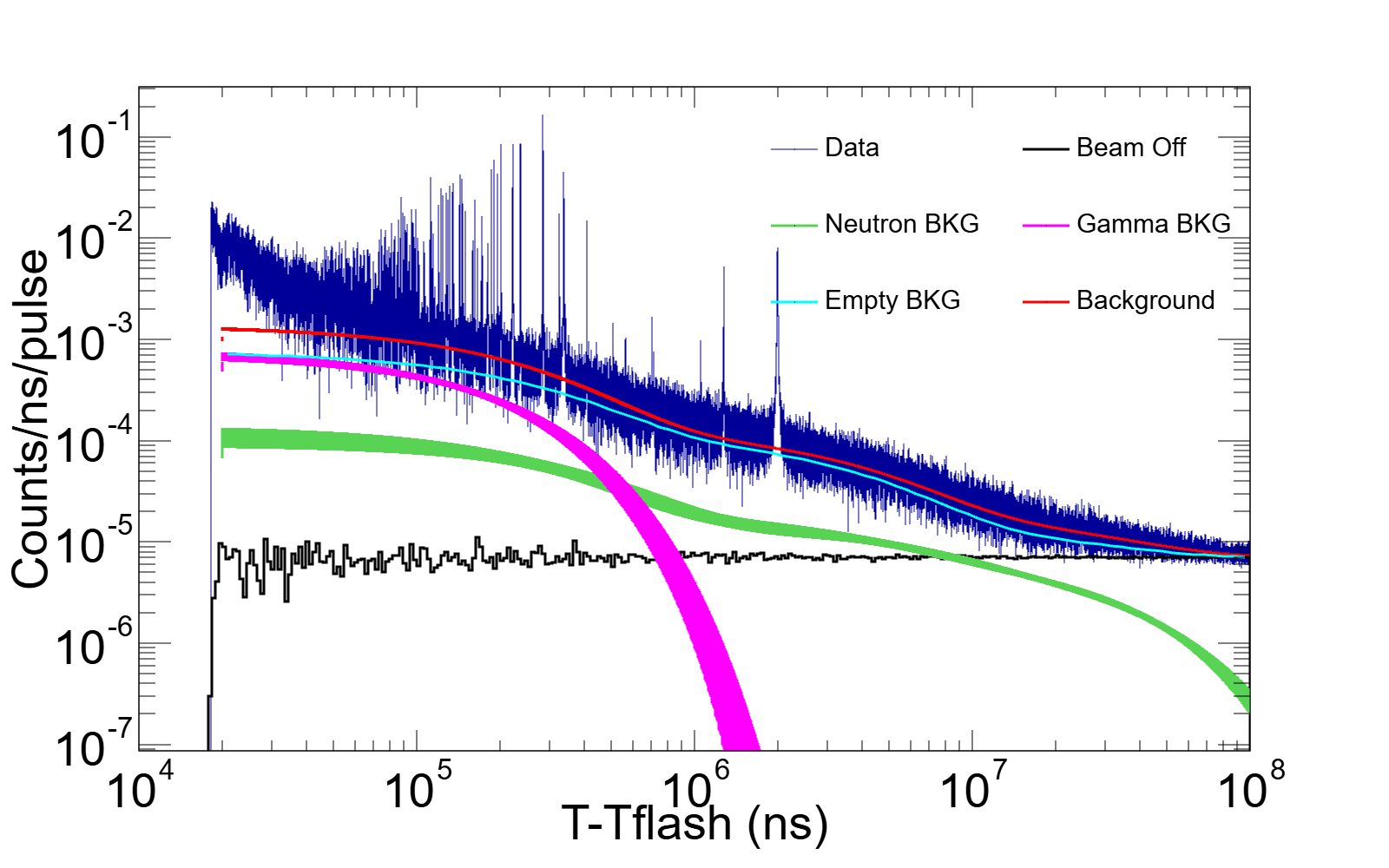}
    \caption{(Color online) Weighted C$_6$D$_6$ time-of-flight spectrum
of $^{94}$Mo sample, together with background components and their uncertainties.The region of interest is between $2.5\times10^4$ ns and $3.0\times10^6$ ns, corresponding to neutron kinetic energy of 270 keV and 20 eV, respectively.}\label{fig:BKG}
\end{figure}

The time-of-flight spectra of $^{94}$Mo together with the different background components and their uncertainties is shown in Fig. \ref{fig:BKG}. The residual background still present in the capture yield after this procedure was estimated and corrected using SAMMY during the resonance fitting.

\subsection{Quality assessment and discussion on uncertainties}

\begin{figure}[]
    \centering
    \includegraphics[width=\columnwidth]{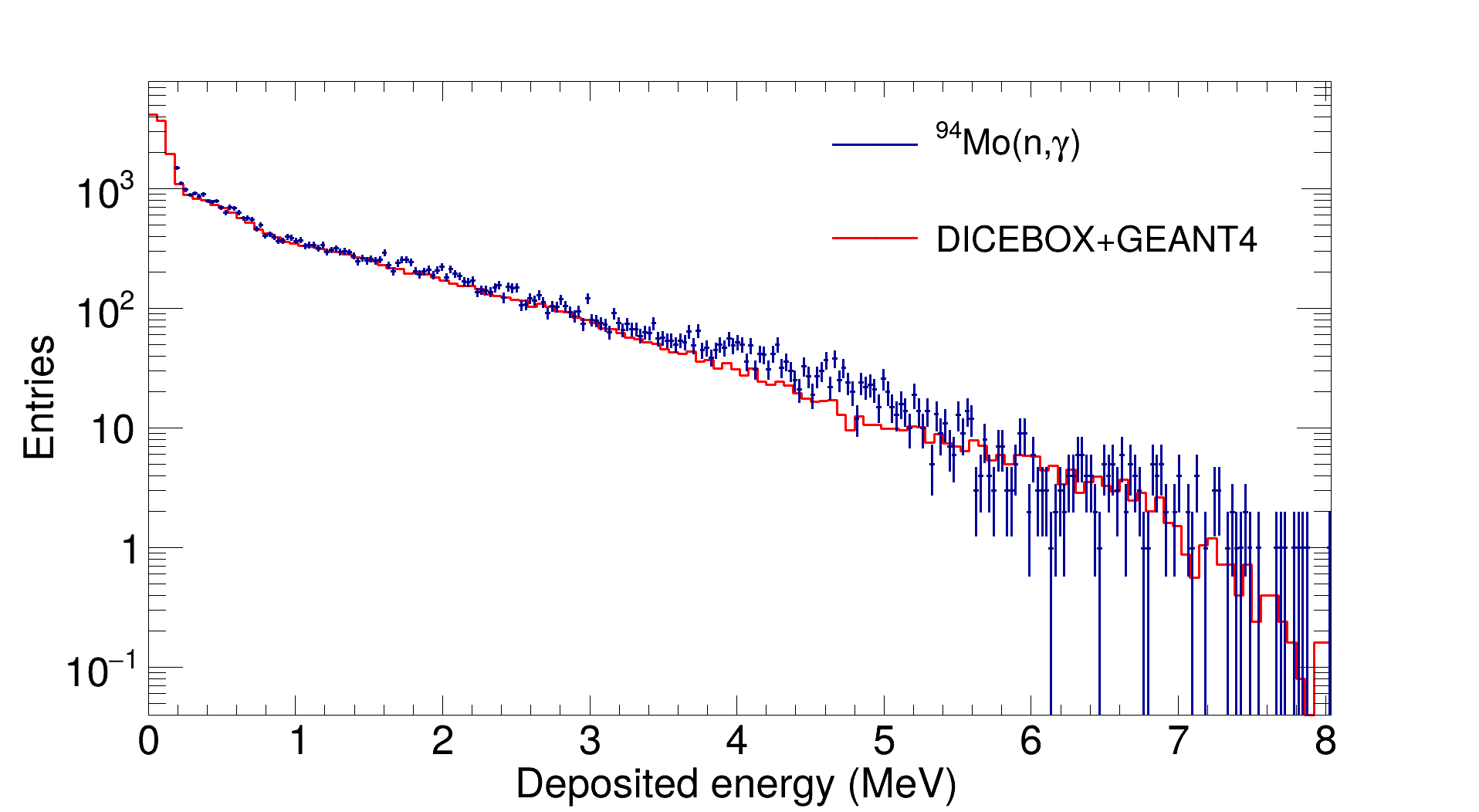}
    \caption{(Color online) Comparison of the experimental spectra of deposited energy in the 1540~eV resonance of $^{94}$Mo (blue) compared with the expected spectrum simulated with DICEBOX and convoluted with the detector resolution using Geant4 (red).}\label{fig:Cascade}
\end{figure}

Particular care was taken in evaluating the systematic uncertainties. The main source of uncertainties comes from the normalization, PHWT, background evaluation and subtraction, sample characteristics and neutron flux. The uncertainty of the normalization factor depends on the difference between the electromagnetic cascade in $^{197}$Au and $^{94}$Mo below the detection thresholds. Two different effects can introduce such uncertainty, low energy photons and conversion electrons. 
To estimate the uncertainty in the normalization given by these effects the technique discussed in Ref.~\cite{Mastromarco19} was applied. The count loss attributable to the detector threshold was estimated by means of Monte Carlo simulations. The $\gamma$ cascades and conversion electrons emitted following the de-excitation of $^{198}$Au and $^{95}$Mo were simulated using the DICEBOX code~\cite{Becvar98}. In DICEBOX experimental data are combined with statistical models to ensure the completeness of the level scheme. The response of the detectors to the simulated cascade was obtained with Geant4 simulations and subsequently convoluted with the experimental resolution estimated using calibration sources. 
The comparison of the simulated and experimental spectrum from one $s$-wave resonance of $^{94}$Mo is shown in Fig.~\ref{fig:Cascade}. The correction due to the weighted count loss below the detector threshold was calculated for both $^{197}$Au and $^{94}$Mo samples, resulting in 3.3\% and 2.7\%, respectively. Therefore, the bias in the normalization due to the lost events, calculated as the difference between the correction factors for Au and $^{94}$Mo, is 0.6\%. The value of the correction deviated by less than 0.5\% for the different detectors and with different simulated spin-parities, therefore this value was used to estimate the total systematic uncertainty in the normalization procedure, together with the positioning error and the error in the area evaluation, leading to a total uncertainty on the normalization of 1.5\%.

The uncertainty on the PHWT was evaluated studying the deviation between the weighted response of the detectors and the total energy of the cascade. Any deviation from unity of this ratio can be considered an uncertainty in the application of the PHWT. The deviation observed for both $^{197}$Au and $^{94}$Mo samples was below 2.0\% for the majority of the energy range, therefore this value has been assumed for the uncertainty on the WF.

The areal density of the sample was obtained with an uncertainty below 0.5\% from a measurement of the weight and the area. The area of the $^{94}$Mo sample was obtained with a microscope system from Mitutoyo~\cite{mitutoyo}. The isotopic enrichment of the $^{94}$Mo sample, as declared by the provider, was verified by comparing the results of its transmission measurement with those obtained from two natural molybdenum samples measured at the 50~m flight path of GELINA. More specifically, the fitted isotopic abundance of the $^{94}$Mo sample was found to be within 1.3\% of the declared value. Therefore, an uncertainty of 1.5\% was assumed for the $^{94}$Mo sample enrichment. Additional transmission measurements using pressed pellets samples with the same dimension of the isotopically enriched one were performed at the 10~m flight path of GELINA to evaluate the effect of grain size of the $^{94}$Mo sample. From a comparison of the transmission obtained with the 5~$\mu$m and the 350~$\mu$m samples, no deviation was observed in the transmission spectra confirming the absence of effects given by inhomogeneity in the powder. The experimental details and results of these additional transmission measurements are reported in \cite{MucciolaThesis}. 
For the estimation of the uncertainty in the evaluated neutron flux, a systematic uncertainty of 2\%, as obtained during the commissioning of the neutron flux, was assumed. 
The uncertainty related to the background subtraction is given by the uncertainty on the evaluated background, as described in Sec.~\ref{Background subtraction}, and the signal-to-background ratio. 
From the results of the background evaluation, an indetermination of the background level below 6.5\% was observed. Considering that the signal-to-background ratio is approximately 2 in the URR and can reach values as high as 100 in the resonance region, this corresponds to a propagated uncertainty in the capture yield---attributable to background subtraction---on the order of 1\% in the RRR and 3\% in the URR. The list of all the uncertainties is reported in Table~\ref{Uncertainties}.

\begin{table}[h] 
\caption{Summary of the correlated uncertainties.}
\begin{tabular}{lcc} \hline \label{Uncertainties}
Source of & Resolved Resonance & Unresolved Resonance\\ 
uncertainty & Region (RRR)  & Region (URR)\\ 

\hline
Normalization & 1.5\% & 1.5\% \\
PHWT & 2\% & 2\% \\
Sample mass & 1.5\% & 1.5\% \\
Neutron flux & 2\% & 2\% \\
Background & 1\% & 3\% \\
\hline
Total & 3.7\% & 4.6\% \\
\hline
\end{tabular}
\end{table}

\section{\label{sec:level1}Transmission measurements}
Transmission experiments using natural and $^{94}$Mo samples were performed at the time-of-flight facility GELINA. 
Details about the accelerator and the neutron-producing target can be found elsewhere~\cite{Mondelaers06,TRONC85}.
The measurements were performed at the 10 m measurement station of flight path 13, which forms an angle of 18$^{\circ}$ with the normal of the face of the moderator view in the flight path.  
The accelerator was operated at 400 Hz and the measurements were carried out with the moderated neutron spectrum. 
A shadow bar made of Cu and Pb was placed close to the uranium target to reduce the intensity of the $\gamma$-ray flash and the fast neutron component. 
The samples and detector were placed in an acclimatized room to keep them at a temperature of 20$^{\circ}$C. The sample was placed at 7.7~m from the moderator while the $^6$Li scintillator was placed in a metallic castle at 10.86~m. The samples were placed on an automatic sample exchanging system controlled by the DAQ.

Two permanent black resonance filters of Na and Co were used to continuously monitor the background at 132~eV and 2850~eV 
and to account for the impact of the sample on the background~\cite{Schillebeeckx12}.
Additional runs with Na (2850~eV), Co (132~eV), W (21.1~eV), and Ag (5.2~eV) black resonance filters were performed to estimate the different background components. 
The neutron beam passing through the sample and filters was collimated to a diameter of about 1 cm and 
detected by a 6.35~mm thick and 101.6~mm diameter NE912 Li-glass scintillator enriched to $95\%$ in $^{6}$Li. 
The scintillator was connected through a boron-free quartz window to a 127 mm EMI 9823 KQB photomultiplier (PMT).

The TOF of a detected neutron was derived from the time difference between the stop signal, obtained from the anode signal of the PMT, and the start signal given at each electron burst. This time difference was processed with a multi-hit fast time coder with a 1~ns time resolution. 
The TOF and the pulse height of each detected event were recorded in list mode using a multi-parameter data acquisition system developed at the JRC Geel (BE). Details about the electronic set-up and data acquisition system are given in Ref.~\cite{Paradela21}. 

The $^{94}$Mo sample used in transmission measurements are the same used in capture measurements at n\_TOF. Two additional natural molybdenum samples were used in the campaign. The natural samples were pressed pellets made of powder with different grain sizes (5 $\mu$m and 350 $\mu$m) with the same 20 mm diameter of the enriched sample. These samples were used to study the effect of the grain size in transmission measurements and to evaluate the isotopic enrichment of the $^{94}$Mo sample. Similarly to the enriched samples, the pressed pellets made from $^{\text{nat}}$Mo powder were put under vacuum in thin plastic bags, the details on these additional transmission measurements are reported in Ref.~\cite{MucciolaThesis}.

The experimental transmission $T_{exp}$ as a function of TOF was obtained from the ratio of a 
sample-in measurement $C_{in}$ and a sample-out measurement $C_{out}$, both corrected for their background contributions $B_{in}$ and $B_{out}$, respectively:

\begin{equation}
\label{trx}
T_{exp}=N_t\frac{C_{in}-KB_{in}}{C_{out}-KB_{out}} \qquad
\end{equation}

The TOF spectra $C_{in}$ and $C_{out}$ were corrected for losses due to the dead time in the detector and electronics chain. The data reduction procedure to obtain $T_{exp}$ has been performed using the AGS (Analysis of Geel Spectra) developed at the IRMM~\cite{Becker12}. This package was used for dead-time correction, background fitting and subtraction, and normalization. The program allows a full propagation of the uncertainty starting from the uncorrelated uncertainty due to counting statistics. The experimental transmission obtained with AGS includes a complete covariance matrix with correlated and uncorrelated uncertainty. 
Since both the neutron flux intensity and energy profile vary during the campaign, the sample-in and sample-out measurements have been divided into short duration cycles of around 600 s alternating between the different configurations. This procedure reduces the uncertainty on the normalization to less than 0.25\%~\cite{Sirakov13}. A normalization factor of $N_T=1.0000 (25)$ was therefore included in Eq.~\ref{trx} to account for this uncertainty. All the spectra were normalized to the same bin-width and the neutron intensity in the target hall. The latter was measured using the BF$_3$ monitors placed in different positions in the target hall. Only cycles for which the ratio between the total counts in the transmission detector and in the neutron monitor deviated by less than 1 $\%$ were selected. The dead time of the detection chain $t_{d}$ = 3305(25) ns was derived from a spectrum of the time interval between successive events. The maximum dead time correction was less than 20$\%$. In Ref.~\cite{Schillebeeckx12}, it has been demonstrated that uncertainties due to such dead time corrections are very small and can be neglected. The flight-path length of $L=10.860 (1)$ m was obtained from previous measurements with $^{238}$U using the reference resonance at 6.673~eV~\cite{Derrien05}. 

The background as a function of TOF was approximated by an analytic expression applying the black resonance technique~\cite{Schillebeeckx12}. 
The factor K = 1.00 (3) in Eq.~\ref{trx} was introduced to account for systematic effects due to the background model. 
Its uncertainty was derived from a statistical analysis of the difference between the observed black resonance dips and the estimated background~\cite{Sirakov13}. 
This uncertainty is only valid for measurements with at least one fixed black resonance filters placed in the beam~\cite{Schillebeeckx12}. 

The analytical function used for approximating the background was a sum of a time-independent and three time-dependent components:

\begin{equation}
\label{bkg}
B(t)=B_0+B_\gamma(t)+B_n(t)+B_{\tau_0}(t) \qquad
\end{equation}
 
\begin{figure}[]
    \centering
    \includegraphics[width=\columnwidth]{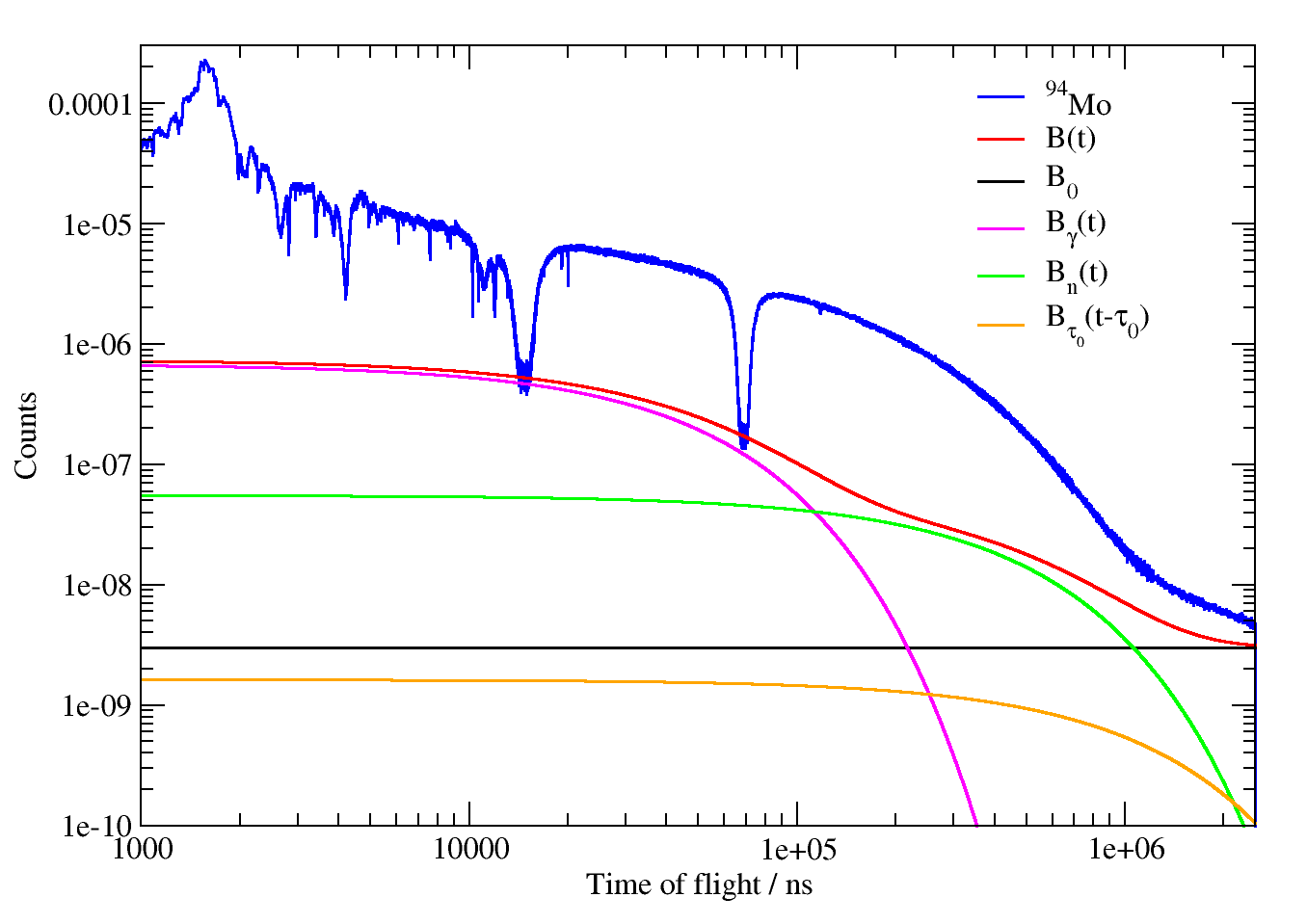}
    \caption{(Color online) Dead-time corrected TOF-spectra resulting from transmission measurements with the $^{94}$Mo sample-in measurement at the 10 m station of GELINA. The total background together with the time independent and time dependent background components are shown. The region of interest is between $5\times10^3$ ns and $2.0\times10^5$ ns, corresponding to neutron kinetic energy of 30 keV and 20 eV, respectively.}\label{fig:BKG_GELINA}
\end{figure}

The time-independent component $B_0$ is related to the ambient background radiation and background contributions that lost any time correlation. 
The first  term is due to the detection of 2.2 MeV $\gamma$-rays resulting from neutron capture in hydrogen present in the moderator. 
The second term originates predominantly from neutrons scattered inside the detector station. 
The third one is attributable to slow neutrons from previous accelerator cycles. 
This contribution was estimated by an extrapolation of the TOF-spectrum at the end of the cycle. 
The time shift $\tau_0$ is the inverse of the accelerator frequency, i.e. $\tau_0$ = 2.5 ms for 400 Hz. 
The free parameters were derived from the results of measurements with all the background filters in the beam. The amplitude of the background was adjusted to the black resonance dips at 132 eV and 2850 eV due to the presence of the Co and Na filters, respectively. The dead-time corrected sample-in TOF-spectrum together with the background
contributions resulting from the measurements of the $^{94}$Mo sample and fixed Co and Na black resonance filters in the
beam are shown in Fig. \ref{fig:BKG_GELINA}.

\section{\label{sec:CrossSection} Cross section analysis}
The cross section for $^{94}$Mo was obtained as a function of resonance parameter for neutron energies below 76~keV using the \textit{R}-matrix code SAMMY. From the resonance parameters obtained in this region, the Average Resonance Parameters (ARP) for s- and p-wave resonances have been extracted. The cross section for energies above 76~keV was obtained using the URR formalism and it was compared with statistical model calculations performed with the ARP obtained in this work.

\subsection{\label{sec:RSA}RRR: Combined Resonance shape analysis}
The present capture data from n\_TOF and the transmission data from GELINA were analyzed using the \textit{R}-matrix code SAMMY~\cite{SAMMY}. Experimental effects due to neutron multiple interactions (also referred to as multiple scattering) in the sample, self-shielding, Doppler broadening, and experimental resolution are properly taken into account within the SAMMY code. 

\begin{figure}[]
    \centering
    \includegraphics[width=\columnwidth]{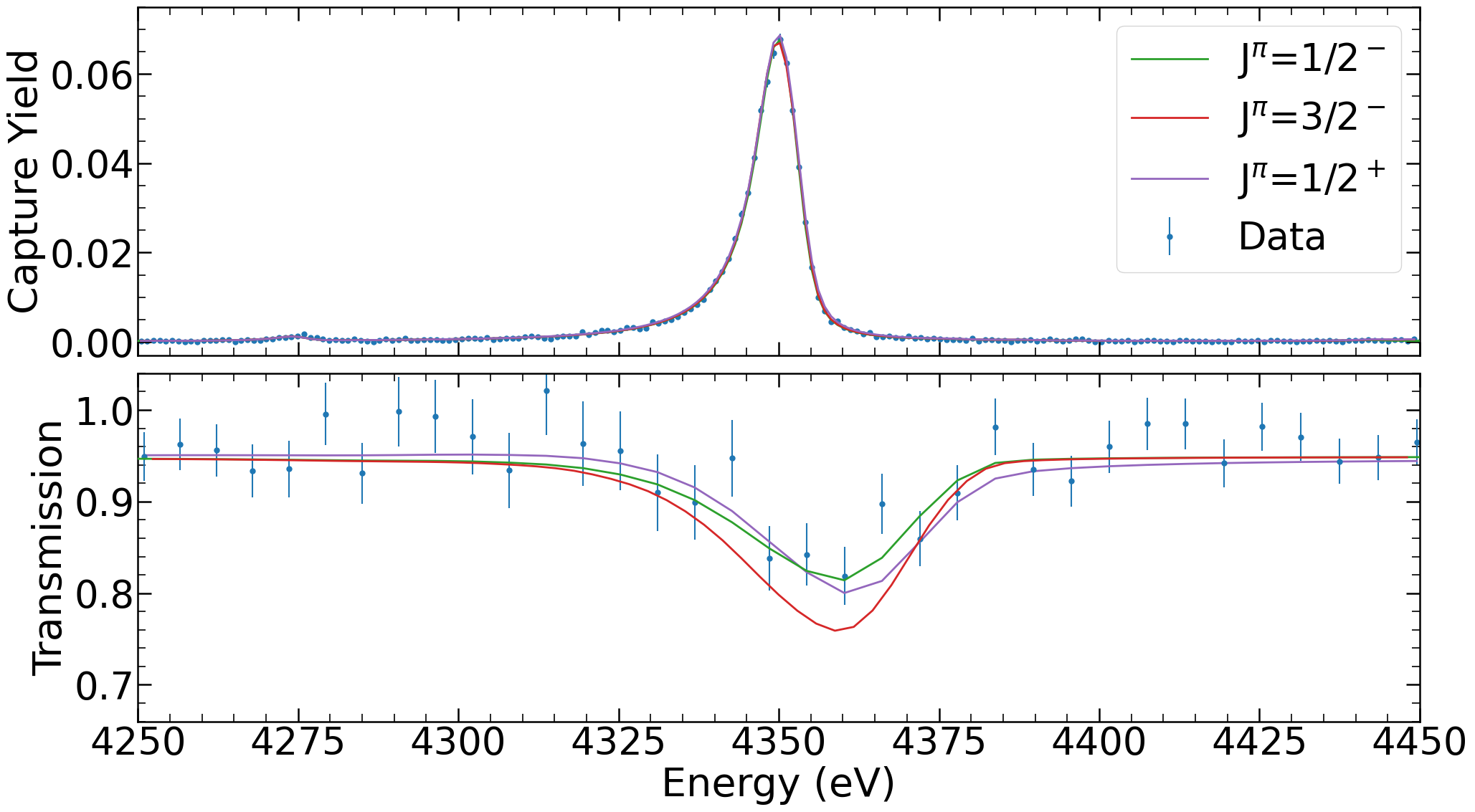}
    \caption{(Color online) Combined resonance shape analysis of capture data from n\_TOF and transmission data from GELINA for the 4357~eV resonance. Results of simultaneous fit using  $J^\pi=0.5^+$ (purple), $J^\pi=0.5^-$ (green), or $J^\pi=1.5^-$ (red) are reported.}\label{fig:SpinCompare}
\end{figure}

The resonance parameters reported in Ref.~\cite{Mucciola22} were used as initial values for the resonance shape analysis. The resonance energies $E_0$, as well as the partial widths $\Gamma_\gamma$ and $\Gamma_n$, were determined through a combined resonance shape analysis (RSA) in the energy region $E_n \le 32$~keV. In the energy region between 32~keV and 76~keV only the capture yield was available, therefore only the capture kernel was obtained from the fit. A total of 186 resonances were observed in the energy region $E_n \le 76$~keV, the 127 resonances above 21~keV are reported here for the first time. 
The spin and parity $J^\pi$ ($0.5^+$ for s-wave resonances, and either $0.5^-$ or $1.5^-$ for p-waves) was kept fixed using values reported in the literature, in particular for the s-wave spin assignments obtained in~\cite{Sheets07}, except for a few cases where the simultaneous fit was not satisfactory using the suggested spin-parity assignment. One example of such simultaneous fit for the 4357 eV resonance is given in Fig.~\ref{fig:SpinCompare}. 

For neutron energies above 21 keV resonance parameters were not available in the literature. In these cases, when both capture and transmission data were available, $J^\pi$ values were assigned based on $\chi^2$ minimization and on the fitted $\Gamma_\gamma$ values, which are expected to be concentrated around the average value derived from data at lower neutron energies and have a maximum value about two times the average i.e. 300-350~meV, as evaluated from statistical models calculations performed with DICEBOX code. 

\begin{figure}[]
    \centering
    \includegraphics[width=\columnwidth]{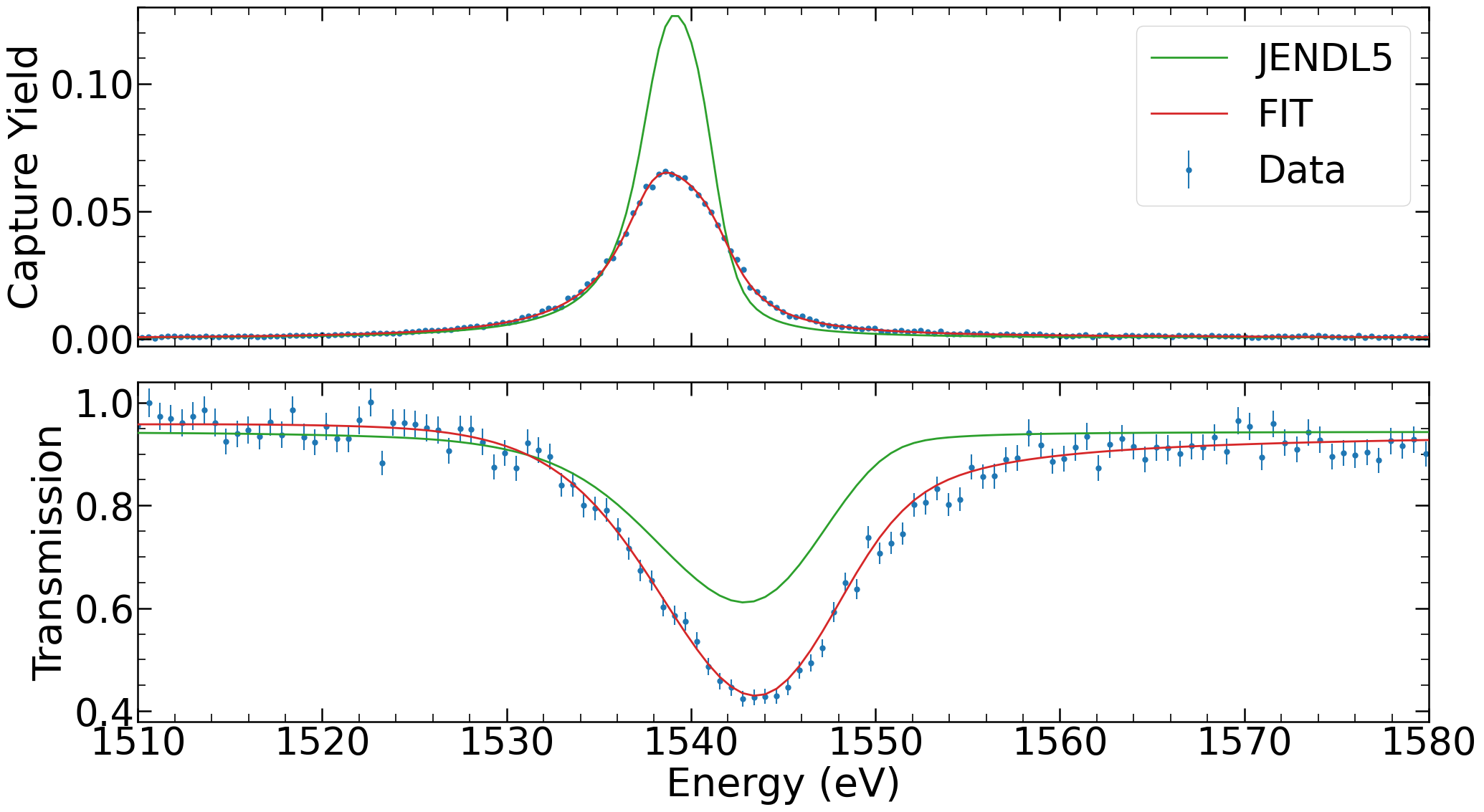}
    \includegraphics[width=\columnwidth]{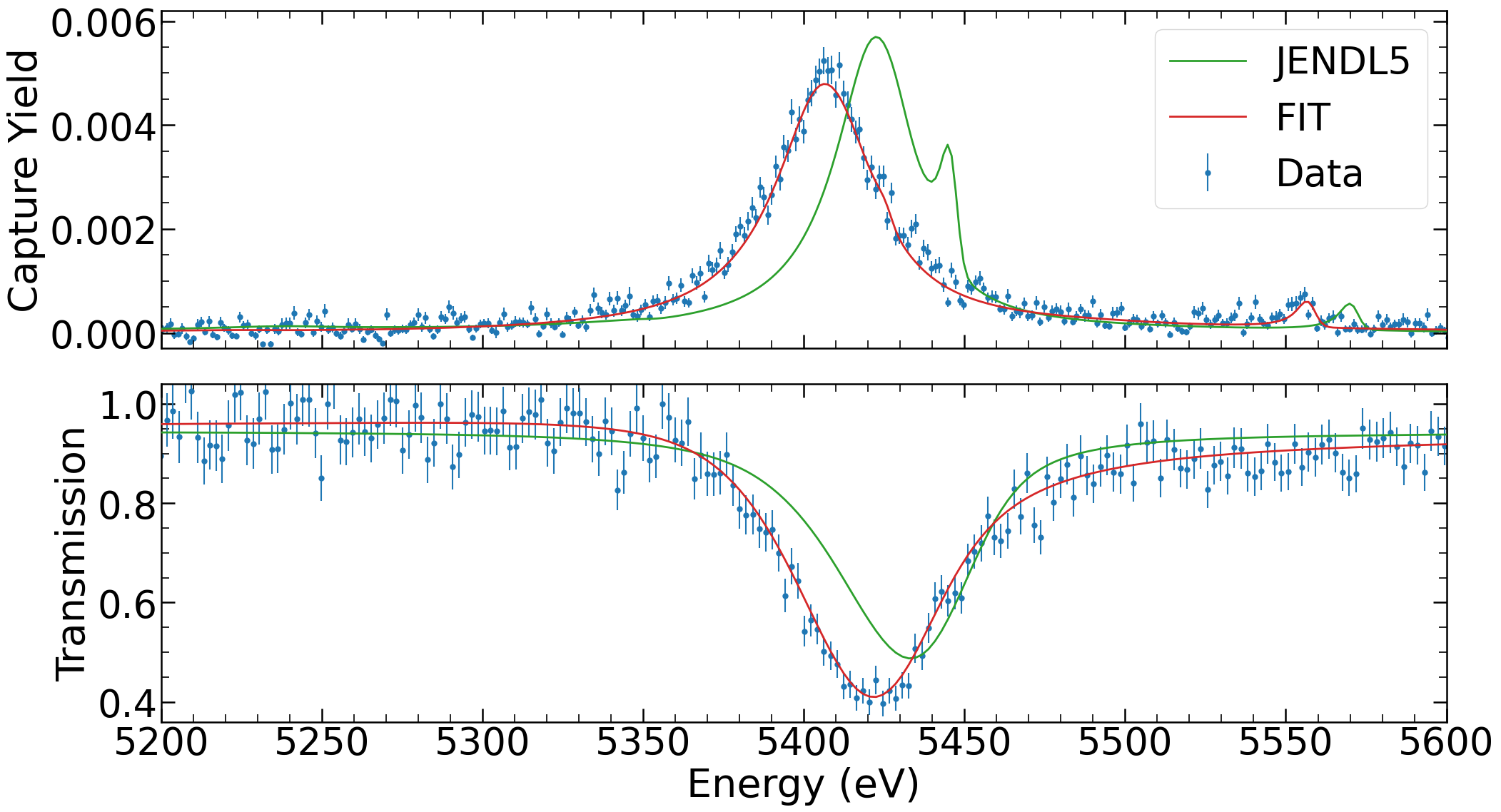}
    \caption{(Color online) Combined resonance shape analysis of capture data from n\_TOF and transmission data from GELINA in the keV region. Expected curves based on resonance parameters available in the JENDL-5 evaluation are reported for comparison.}\label{fig:RSA1}
\end{figure}

\begin{figure}[]
    \centering
    \includegraphics[width=\columnwidth]{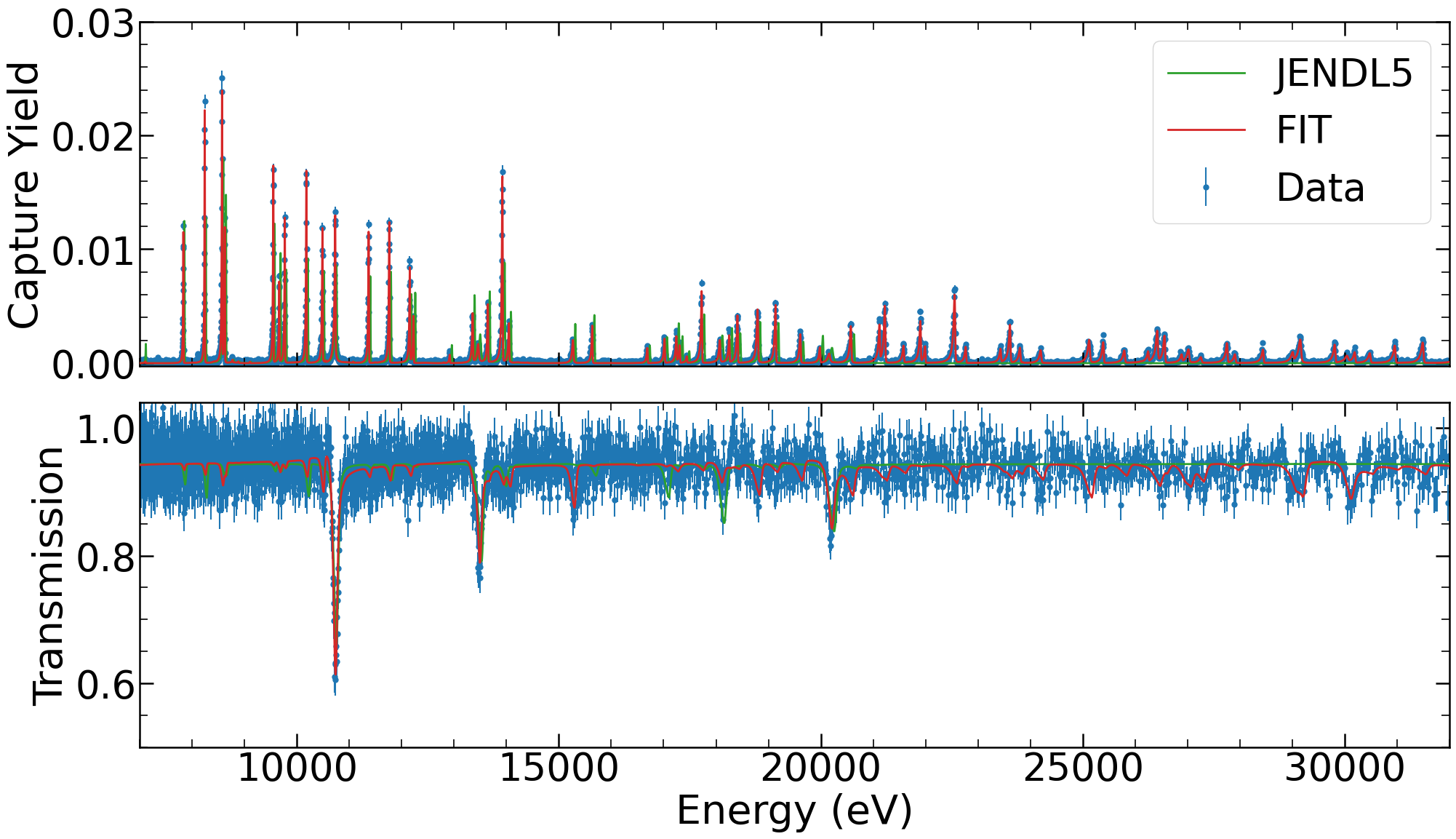}
    \caption{(Color online) Combined resonance shape analysis of capture data from n\_TOF and transmission data from GELINA in the tens of keV region. Expected curves based on resonance parameters available in the JENDL-5 evaluation are reported for comparison.}\label{fig:RSA2}
\end{figure}

\begin{figure}[]
    \centering
    \includegraphics[width=\columnwidth]{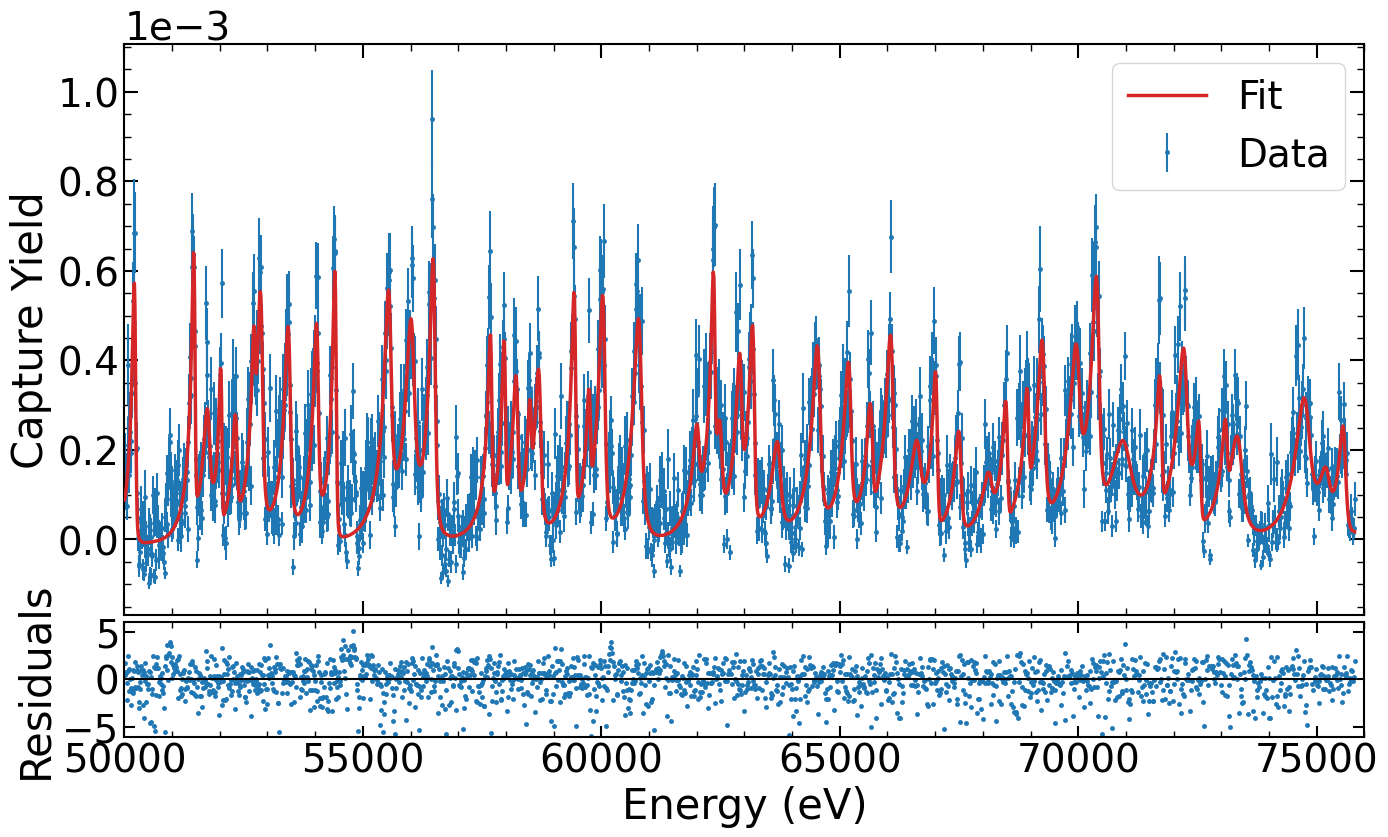}
    \caption{(Color online) Resonance shape analysis of capture data from n\_TOF above 50~keV. The residuals are shown in the bottom panel.}\label{fig:RSA3}
\end{figure}

In Table~\ref{tab:res} the results are reported together with the uncertainties from the fitting procedure. Individual parameters obtained from the combined fit of capture and transmission are reported if the resulting uncertainty was below 50~\%. Other examples of the combined RSA fits are given in Fig.~\ref{fig:RSA1} and Fig.~\ref{fig:RSA2}. The fitted resonances in the energy region above 50~keV, where only capture data is available, are shown in Fig~\ref{fig:RSA3}.
Compared to evaluations, one neutron resonance had to be removed in the energy range covered in Table~\ref{tab:res}. This weak resonance at 7.1 keV was assigned in previous molybdenum experiments, but not observed in this work. The thermal cross section of $^{94}$Mo is strongly dominated by bound resonances. For completeness, a bound resonance with parameters from~\cite{Mughabghab18} was added to Table~\ref{tab:res}.

\setlength{\LTcapwidth}{\columnwidth}
\begin{longtable}[H]{lccccc}

\caption{$\text{n} + {}^{94}$Mo resonance parameters extracted from the simultaneous R-matrix analysis. The quoted uncertainties are from the fit. Spin and parity are taken from Ref.~\cite{Mucciola22}. The column with $\rho(\Gamma_\gamma,\Gamma_n)$ reports the correlation between partial widths.}\\
\label{tab:res}\\

\hline\hline
$E_0$ & $J^\pi$ & $\Gamma_\gamma$  & $g\Gamma_n$ & $\rho(\Gamma_\gamma,\Gamma_n)$ & ${g\Gamma_\gamma\Gamma_n}/\Gamma$   \\
($\mathrm{eV}$) & & ($\mathrm{meV}$) & ($\mathrm{meV}$) & (\%) & ($\mathrm{meV}$) \\
\hline
\endfirsthead

\caption[]{(Continued)}\\
\hline\hline
$E_0$ & $J^\pi$  & $\Gamma_\gamma$  & $g\Gamma_n$ & $\rho(\Gamma_\gamma,\Gamma_n)$ & ${g\Gamma_\gamma\Gamma_n}/\Gamma$   \\
($\mathrm{eV}$) & & ($\mathrm{meV}$) & ($\mathrm{meV}$) & (\%) & ($\mathrm{meV}$) \\
\hline
\endhead

\hline
\endfoot

\hline\hline
\multicolumn{6}{l}{$^a$) $J^\pi$ changed to perform simultaneous fit} \\
\multicolumn{6}{l}{$^b$) Assumed $g=1$ from $g\Gamma_\gamma$} \\
\multicolumn{6}{l}{$^c$) Assumed $g=2$ from $g\Gamma_\gamma$} \\
\multicolumn{6}{l}{$^d$) Firm s-wave by shape or from \cite{Sheets07}} \\
\multicolumn{6}{l}{$^e$) Firm p-wave by shape or from \cite{Sheets07}} \\
\endlastfoot

-143	&	$0.5^+$	&	128	&	148.8	&		&		\\
108.714(1)	&	$0.5^-$	&	146(3)	&	0.1715(7)	&	23	&	0.1713(7)	\\
1051.481(9)	&	$1.5^-$	&	220(20)	&	4.57(4)		&	13	&	4.53(4)	\\
1541.91(1)$^d$	&	$0.5^+$	&	123.3(4)&	1672(7)		&	-13	&	114.8(3)	\\
1656.44(1)	&	$1.5^-$	&	170(30)	&	9.20(9)		&	-24	&	8.95(8)	\\
2174.113(8)	&	$1.5^-$	&	184(2)	&	440(10)		&	-88	&	199(1)	\\
3155.10(1)	&	$1.5^-$	&	96(2)	&	380(30)		&	-93	&	128(1)	\\
3565.92(2)$^a$	&	$0.5^-$	&	250(20)	&	260(30)		&	-99	&	129(2)	\\
3593.55(3)$^d$	&	$0.5^+$	&	116.6(8)&	2750(30)	&	-8	&	111.8(7)	\\
4357.31(3)$^{a,e}$	&	$0.5^-$	&	333(2)	&	1700(30)	&	-99	&	278.7(7)	\\
4610.47(2)$^d$	&	$0.5^+$	&	122(1)	&	510(30)		&	-62	&	98.5(9)	\\
4915.18(3)$^a$	&	$1.5^-$	&	97(3)	&	470(60)		&	-17	&	138(6)	\\
5122.14(3)$^a$	&	$0.5^-$	&	166(2)	&	820(40)		&	-35	&	138(1)	\\
5418.6(2)$^d$	&	$0.5^+$	&	120(1)	&	20500(200)	&	-6	&	119(1)	\\
5434(1)	    &	$0.5^-$		&		&	1.2(5)		&	2	&	1.1(5)	\\
5937.62(6)	&	$0.5^-$	&		&	70(20)		&	-99	&	52(6)	\\
6015.8(2)	&	$0.5^-$	&		&	8.4(6)		&	-74	&	7.8(4)	\\
6392.5(1)	&	$0.5^-$	&		&	25(5)		&	-98	&	23(2)	\\
6700.96(5)$^d$	&	$0.5^+$	&	116(1)	&	1510(80)	&	-2	&	108(1)	\\
6942.7(1)	&	$1.5^-$	&		&	30(3)		&	-95	&	27.9(7)	\\
7844.83(6)$^d$	&	$0.5^+$	&	120(10)	&	300(100)	&	-99	&	82(2)	\\
8254.25(5)	&	$1.5^-$	&	130(6)	&	600(100)	&	-96	&	181(6)	\\
8585.45(6)	&	$1.5^-$	&	147(4)	&	1100(200)	&	-87	&	231(4)	\\
8636.88(8)	&	$0.5^-$	&	135(3)	&	800(100)	&	-69	&	115(2)	\\
9565.21(7)	&	$1.5^-$	&	133(9)	&	600(200)	&	-97	&	187(8)	\\
9678.5(1)$^d$	&	$0.5^+$	&	92(2)	&	700(200)	&	-66	&	82(2)	\\
9785.21(9)$^b$	&	$0.5^-$	&	200(20)	&	500(100)	&	-98	&	145(3)	\\
10197.35(8)	&	$1.5^-$	&	136(2)	&	1500(200)	&	-57	&	229(4)	\\
10504.8(1)	&	$1.5^-$	&	110(2)	&	3600(300)	&	44	&	208(3)	\\
10732.0(8)	&	$0.5^+$	&	78(4)	&	31700(800)	&	-42	&	78(4)	\\
10748.4(1)	&	$1.5^-$	&	166(4)	&	5100(400)	&	57	&	311(7)	\\
11388.7(1)	&	$1.5^-$	&	120(10)	&	900(300)	&	-98	&	190(3)	\\
11782.2(1)	&	$1.5^-$	&	134(2)	&	1600(300)	&	-59	&	230(4)	\\
12175.2(1)	&	$0.5^-$	&	187(4)	&	1300(200)	&	-61	&	163(3)	\\
12245.8(2)	&	$0.5^-$	&		&		&	-99	&	70(10)	\\
12938.2(6)	&	$0.5^-$	&		&	16(2)		&	-71	&	14(1)	\\
13377.4(2)	&	$0.5^-$	&	126(3)	&	3400(400)	&	43	&	122(3)	\\
13484.7(5)	&	$0.5^+$	&	94(2)	&	19300(800)	&	20	&	94(2)	\\
13667.0(2)$^b$	&	$0.5^-$	&	142(4)	&	1000(300)	&	-66	&	125(3)	\\
13945.4(1)	&	$1.5^-$	&	278(6)	&	2200(300)	&	-78	&	445(8)	\\
14070.6(2)	&	$0.5^-$	&	109(2)	&	3100(400)	&	14	&	105(2)	\\
15296.1(4)	&	$0.5^+$	&	106(3)	&	9900(800)	&	39	&	105(3)	\\
15663.4(3)	&	$0.5^-$	&		&		&	-99	&	100(10)	\\
16716.1(6)	&	$0.5^-$	&		&	60(20)		&	-98	&	44(3)	\\
17045.8(4)	&	$0.5^+$	&	90(30)	&		&	-99	&	80(10)	\\
17272.6(4)	&	$0.5^-$	&	98(3)	&	1500(700)	&	9	&	92(4)	\\
17335.4(4)	&	$1.5^-$	&		&	40(10)		&	-98	&	57(3)	\\
17471(1)	&	$0.5^-$	&		&	24(3)		&	-77	&	21(2)	\\
17756.6(2)	&	$1.5^-$	&	158(7)	&	1800(700)	&	-87	&	269(8)	\\
18103.4(5)	&	$0.5^+$	&	105(3)	&	4900(800)	&	11	&	103(3)	\\
18282.6(4)	&	$0.5^-$	&	130(30)	&		&	-98	&	100(4)	\\
18440.9(2)	&	$1.5^-$	&	110(10)	&		&	-97	&	185(8)	\\
18826.7(3)	&	$1.5^-$	&	158(3)	&	10300(900)	&	23	&	306(6)	\\
19169.3(2)	&	$1.5^-$	&	146(4)	&	2600(800)	&	-72	&	262(6)	\\
19642.4(4)	&	$0.5^-$	&	170(4)	&	5000(700)	&	12	&	165(4)	\\
20016.5(6)	&	$0.5^-$	&	77(7)	&		&	-89	&	70(3)	\\
20199(1)	&	$0.5^+$	&	98(4)	&	27000(2000)	&	16	&	97(4)	\\
20610.0(4)$^c$	&	$1.5^-$	&	133(3)	&	11000(1000)	&	35	&	259(7)	\\
21155.6(4)	&	$0.5^-$	&	230(20)	&	1200(700)	&	-93	&	191(7)	\\
21259.7(3)	&	$1.5^-$	&	196(4)	&	5400(900)	&	18	&	366(8)	\\
21615.4(6)	&	$0.5^-$	&	105(4)	&	3000(1000)	&	-11	&	101(4)	\\
21938.1(4)	&	$1.5^-$	&	163(4)	&	1200(800)	&	-23	&	260(40)	\\
22027.7(5)	&	$0.5^-$	&		&			&	-99	&	100(10)	\\
22594.0(3)	&	$1.5^-$	&	288(5)	&	8000(1000)	&	7	&	540(10)	\\
22805.1(6)	&	$0.5^-$	&	114(8)	&			&	-83	&	102(5)	\\
23470.7(8)	&	$0.5^+$	&	94(5)	&			&	-54	&	89(4)	\\
23652.0(4)	&	$1.5^-$	&	159(4)	&	5000(1000)	&	28	&	300(9)	\\
23843.1(6)	&	$0.5^-$	&	126(4)	&	4000(1000)	&	32	&	123(5)	\\
24240.2(7)	&	$0.5^-$	&	120(5)	&	8000(2000)	&	-3	&	119(5)	\\
25168.8(6)	&	$1.5^-$	&	142(4)	&	20000(3000)	&	34	&	279(7)	\\
25435(1)	&	$0.5^-$	&	196(9)	&			&	-70	&	178(7)	\\
25839.0(9)	&	$0.5^-$	&	132(5)	&	6000(2000)	&	4	&	129(5)	\\
26293.6(9)	&	$0.5^-$	&	140(60)	&			&	-98	&	84(7)	\\
26469.0(6)	&	$1.5^-$	&	181(5)	&	11000(2000)	&	38	&	350(10)	\\
26607.0(5)	&	$1.5^-$	&	145(4)	&	5000(2000)	&	27	&	280(10)	\\
26936(1)	&	$0.5^+$	&	73(5)	&			&	15	&	71(5)	\\
27067(1)	&	$0.5^-$	&	178(7)	&	13000(2000)	&	13	&	175(6)	\\
27304(2)	&	$0.5^+$	&	75(4)	&	13000(3000)	&	-14	&	74(4)	\\
27802.0(7)	&	$0.5^-$	&	280(70)	&			&	-99	&	200(20)	\\
27955(1)	&	$0.5^+$	&	99(5)	&		&	28	&	96(5)	\\
28490.0(8)	&	$0.5^-$	&	170(30)	&			&	-95	&	140(10)	\\
29060(1)	&	$0.5^+$	&	112(6)	&	6000(2000)	&	-27	&	106(6)	\\
29207.2(7)	&	$1.5^-$	&	215(5)	&	27000(3000)	&	33	&	420(10)	\\
29863.6(7)	&	$1.5^-$	&		&			&	-99	&	210(10)	\\
30104(2)	&	$0.5^+$	&	138(7)	&	30000(4000)	&	18	&	138(7)	\\
30245(1)	&	$1.5^-$	&		&	190(80)		&	-98	&	128(8)	\\
30535(1)	&	$0.5^-$	&	142(6)	&	5000(2000)	&	10	&	138(6)	\\
31012.7(7)	&	$1.5^-$	&	132(8)	&			&	-84	&	240(10)	\\
31547.4(7)	&	$1.5^-$	&	166(5)	&	7000(3000)	&	36	&	320(10)	\\
32259.8(7)	&		&		&			&		&	320(10)	\\
32708.7(9)	&		&		&			&		&	225(9)	\\
33111(2)	&		&		&			&		&	70(6)	\\
33410(1)	&		&		&			&		&	86(7)	\\
33796(4)	&		&		&			&		&	89(8)	\\
34375(2)	&		&		&			&		&	370(20)	\\
34766(2)	&		&		&			&		&	157(10)	\\
35313(4)	&		&		&			&		&	140(10)	\\
35734(3)	&		&		&			&		&	100(10)	\\
35864(2)	&		&		&			&		&	229(20)	\\
35936(2)	&		&		&			&		&	224(20)	\\
37211(2)	&		&		&			&		&	85(9)	\\
37809(2)	&		&		&			&		&	191(20)	\\
37895(2)	&		&		&			&		&	390(20)	\\
38376(2)	&		&		&			&		&	290(20)	\\
38493(2)	&		&		&			&		&	135(10)	\\
39045(2)	&		&		&			&		&	360(20)	\\
39171(2)	&		&		&			&		&	420(20)	\\
39391(2)	&		&		&			&		&	310(10)	\\
40116(2)	&		&		&			&		&	138(9)	\\
40898(4)	&		&		&			&		&	65(8)	\\
41238(5)	&		&		&			&		&	190(20)	\\
41338(2)	&		&		&			&		&	230(20)	\\
41873(2)	&		&		&			&		&	490(20)	\\
42219(2)	&		&		&			&		&	310(10)	\\
42398(4)	&		&		&			&		&	71(8)	\\
42783(1)	&		&		&			&		&	260(10)	\\
43201(3)	&		&		&			&		&	140(10)	\\
43319(3)	&		&		&			&		&	90(10)	\\
43622(2)	&		&		&			&		&	370(10)	\\
44167(2)	&		&		&			&		&	560(20)	\\
44971(2)	&		&		&			&		&	240(20)	\\
45071(4)	&		&		&			&		&	150(10)	\\
45304(3)	&		&		&			&		&	130(10)	\\
45925(2)	&		&		&			&		&	310(20)	\\
46304(2)	&		&		&			&		&	280(20)	\\
46798(2)	&		&		&			&		&	410(20)	\\
47108(2)	&		&		&			&		&	510(20)	\\
47635(4)	&		&		&			&		&	100(10)	\\
47789(3)	&		&		&			&		&	190(20)	\\
47964(3)	&		&		&			&		&	480(20)	\\
48712(2)	&		&		&			&		&	350(70)	\\
48848(3)	&		&		&			&		&	240(10)	\\
49196(2)	&		&		&			&		&	260(20)	\\
49472(5)	&		&		&			&		&	80(20)	\\
49589(3)	&		&		&			&		&	540(20)	\\
50306(2)	&		&		&			&		&	330(20)	\\
51552(2)	&		&		&			&		&	330(20)	\\
51848(7)	&		&		&			&		&	270(20)	\\
52119(3)	&		&		&			&		&	170(10)	\\
52432(6)	&		&		&			&		&	180(10)	\\
52812(6)	&		&		&			&		&	150(20)	\\
52962(4)	&		&		&			&		&	500(30)	\\
53547(4)	&		&		&			&		&	440(20)	\\
54139(4)	&		&		&			&		&	380(20)	\\
54519(2)	&		&		&			&		&	400(20)	\\
55655(5)	&		&		&			&		&	580(30)	\\
56139(5)	&		&		&			&		&	640(30)	\\
56595(3)	&		&		&			&		&	610(20)	\\
57786(3)	&		&		&			&		&	290(20)	\\
58072(3)	&		&		&			&		&	250(20)	\\
58329(5)	&		&		&			&		&	310(20)	\\
58613(6)	&		&		&			&		&	130(20)	\\
58809(4)	&		&		&			&		&	420(20)	\\
59546(3)	&		&		&			&		&	410(20)	\\
59857(5)	&		&		&			&		&	170(20)	\\
60155(4)	&		&		&			&		&	610(30)	\\
60919(5)	&		&		&			&		&	730(30)	\\
62131(10)	&		&		&			&		&	210(20)	\\
62471(4)	&		&		&			&		&	360(30)	\\
62607(10)	&		&		&			&		&	280(40)	\\
63029(6)	&		&		&			&		&	372(30)	\\
63299(4)	&		&		&			&		&	500(30)	\\
63834(9)	&		&		&			&		&	320(20)	\\
64674(5)	&		&		&			&		&	620(30)	\\
65323(5)	&		&		&			&		&	510(30)	\\
65741(10)	&		&		&			&		&	70(40)	\\
65788(7)	&		&		&			&		&	200(20)	\\
66207(5)	&		&		&			&		&	640(30)	\\
66768(10)	&		&		&			&		&	340(30)	\\
67143(4)	&		&		&			&		&	380(30)	\\
67639(6)	&		&		&			&		&	310(20)	\\
68269(10)	&		&		&			&		&	190(20)	\\
68616(4)	&		&		&			&		&	270(20)	\\
69062(5)	&		&		&			&		&	240(20)	\\
69389(5)	&		&		&			&		&	650(30)	\\
70106(8)	&		&		&			&		&	670(40)	\\
70532(5)	&		&		&			&		&	870(50)	\\
71145(10)	&		&		&			&		&	690(30)	\\
71854(5)	&		&		&			&		&	360(30)	\\
72380(7)	&		&		&			&		&	760(40)	\\
72672(5)	&		&		&			&		&	270(30)	\\
73228(7)	&		&		&			&		&	190(30)	\\
73526(10)	&		&		&			&		&	580(30)	\\
74930(10)	&		&		&			&		&	800(50)	\\
75378(20)	&		&		&			&		&	250(40)	\\
75719(6)	&		&		&			&		&	280(30)	\\

\end{longtable}

\subsection{\label{sec:AVERAGE}Statistical properties of resonance parameters}

Resolved resonance parameters obtained in the previous section were used to derive the basic statistical properties of the resonances, following the same prescriptions in Ref.~\cite{Mastromarco19}. 
Parameters of individual resonance below 32 keV can were used to derive the average resonance parameters. During this analysis, we assumed that the parity of all resonances was correctly assigned.

Neutron strength functions $S_0$ and $S_1$ were derived from a sum of reduced neutron widths below 32 keV. In practice, a small correction for missed resonance strength was applied considering the values of average resonance parameters listed in Tab.~\ref{tab:averageparam}. The channel radius of $R= 1.23 A^{1/3} + 0.8 \ \rm{fm} = 6.4$~fm was adopted for determination of $S_1$. 

For determination of average resonance spacings $D_0$ and $D_1$ we assumed that positions of resonances with the same spin and parity obey long-range correlations predicted from the Gaussian Orthogonal Ensemble~\cite{Dyson63}. The values were then obtained from the energy difference between the first and last observed resonance. Similarly to Ref.~\cite{Knapova22} we further applied the correction for possible sub-threshold resonances and assumed that the position of the last resonance can fluctuate.   

The average radiation widths $\langle \Gamma_\gamma\rangle_0$ and $\langle \Gamma_\gamma\rangle_1$ and the widths of $\Gamma_\gamma$ distributions $\sigma_{\Gamma_\gamma 0}$ and $\sigma_{\Gamma_\gamma 0}$ were determined using the 
maximum likelihood fit assuming a Gaussian distribution of individual $\Gamma_\gamma$ values, similarly to Ref.~\cite{Lederer19}. We adopted all the resonances below 32 keV with provided $\Gamma_\gamma$ for this purpose. The larger width of the distribution $\sigma_{\Gamma_\gamma 1}$ than $\sigma_{\Gamma_\gamma 0}$ is expected from DICEBOX simulations, especially for $1/2^-$ resonances. These simulations further expect the average radiation widths for $1/2^-$ to be larger than that for $3/2^-$ by about 10-15~\%.

The resulting values of the average resonance parameters for s- and p-wave resonance are reported in Table~\ref{tab:averageparam} and are compared to the ones reported by Musgrove~\cite{Musgrove76} and the values reported in the work of Mughabghab~\cite{Mughabghab18} and in the RIPL-3 compilation~\cite{Capote09}. 

\begin{table}[h] 
\caption{Average resonance parameters from this work together with the ones from Musgrove~\cite{Musgrove76}, Mughabghab~\cite{Mughabghab18}, and RIPL-3~\cite{Capote09} for comparison.}
\begin{tabular}{lcccc} 
\hline \label{tab:averageparam}
& This work & Ref~\cite{Musgrove76} & Ref~\cite{Mughabghab18} & Ref~\cite{Capote09}\\
\hline
$D_0$ (eV) & $1470(260)$ & $1150(350)$ & 1694(390) & 1320(180)\\ 
$D_1$ (eV)& $460(80)$ & 450(40) & 508(47) \\
$10^{4}\times S_0$ & $0.48(15)$& $ 0.45(25)$ & $0.70(34)$ & 0.44(8)\\
$10^{4}\times S_1$ & $4.8(9)$ & $7.5(2.5)$ & $7.23(2.67)$ & 7.2(2.7)\\
$\langle{\Gamma}_\gamma\rangle_0$ (meV) & $104(4)$ & $135(20)$ & 137(9)  & 135(15)\\
$\langle{\Gamma}_\gamma\rangle_1$ (meV) & $158(7) $  & $175(30) $  & 230(15) & 188(10)\\
$\sigma_{\Gamma_{\gamma0}}$ & 17(3) &&&\\
$\sigma_{\Gamma_{\gamma1}}$ & 49(5) &&&\\
\hline
\end{tabular}
\end{table}

\subsection{\label{sec:URR}URR: Cross section in the continuum region} 
\begin{figure}[]
    \includegraphics[width=\columnwidth]{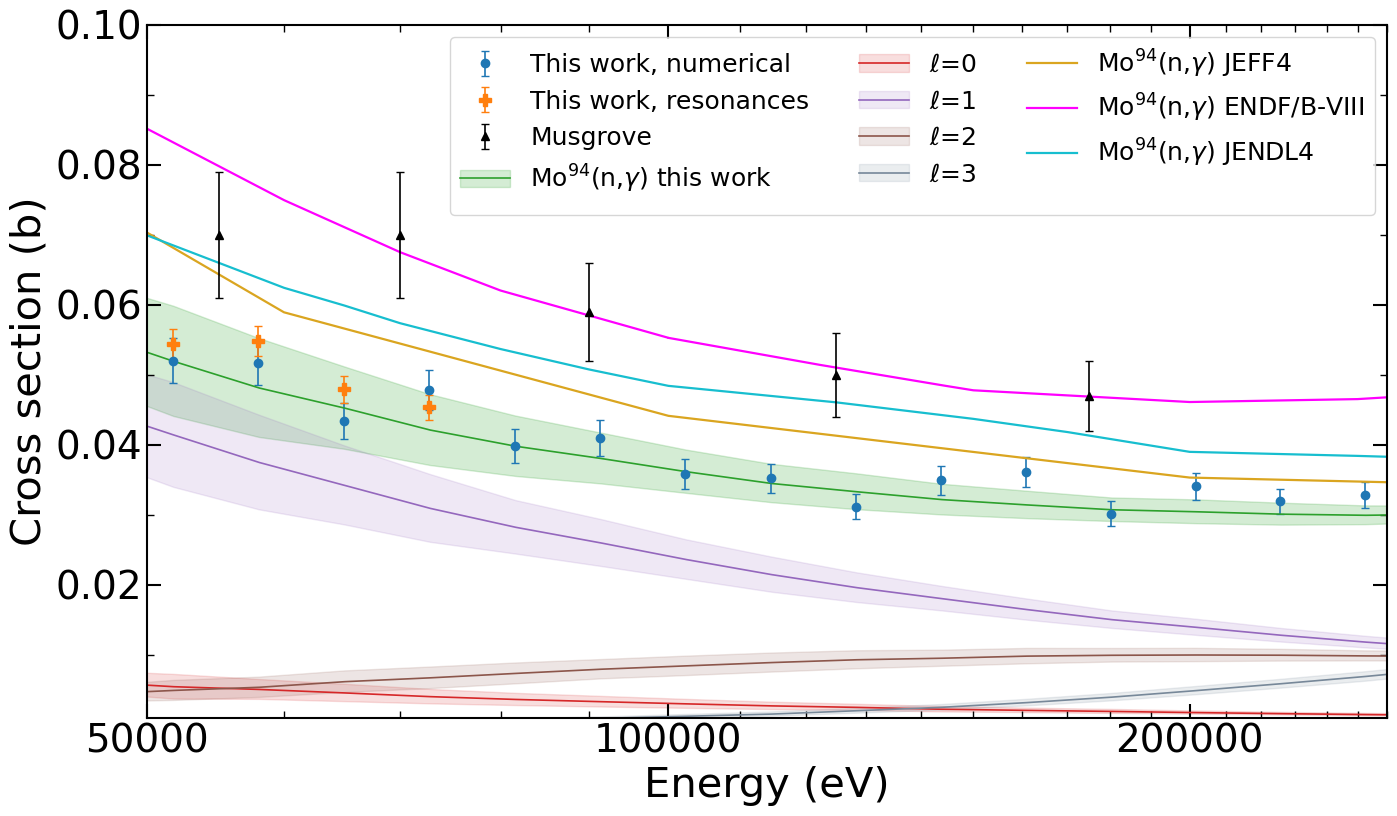}
    \caption{(Color online) $^{94}$Mo(n,$\gamma$) cross section data in the unresolved resonance region, obtained via numerical calculation (blue) and from resonance parameters (orange). The experimental data obtained at n\_TOF are compared to a statistical model calculations based on average resonance parameters from this work. Different contributions of neutrons with different $\ell$ are also reported. The experimental data are also compared with evaluations and experimental data by Musgrove~\cite{Musgrove76}.}\label{fig:URR}
\end{figure}

Above 76 keV the experimental resolution became too low to clearly resolve individual resonance structures and the cross section was obtained using the URR formalism. From the capture yield obtained in the n\_TOF measurement, it was possible to extract the average $^{94}$Mo(n,$\gamma)$ cross section between 50 and 250 keV with a coarse binning of 20 bins per energy decade, as showed in Fig.~\ref{fig:URR}. This cross section is obtained after corrections for multiple scattering and self-shielding effect, estimated using Monte Carlo simulations as described in~\cite{Mingrone17}. The data are also corrected for the effects of the contamination from other molybdenum isotopes, using the isotopic composition in Tab.~\ref{MoEnrichment}. The corrections applied to the cross section are of the order of 4\% in all the energy regions. The uncertainty of these corrections has been estimated from the deviations in the correction obtained in the different detectors, which accounts for an additional 2\% in the uncertainty in the cross section. Between 50 and 76~keV we obtained the cross section both from URR approach and from the individual resonance parameters reported in this work as an average cross section in a bin. The good agreement of the cross section from both approaches can be seen in Fig.~\ref{fig:URR} and it confirms the accuracy of the background subtraction and multiple scattering and self-shielding corrections applied in the cross section determination in the URR formalism.

Fig.~\ref{fig:URR} also shows results of cross section calculations using statistical model. For these calculations we generated individual resonance sequences based on the average resonance parameters considering expected statistical fluctuations of individual quantities. This approach allows us to calculate not only expectation cross section values but also expected fluctuations in individual bins. The colored bands represent standard deviations obtained from 500 simulated sequences. In addition to the total capture cross section, we also present contributions from individual neutron orbital momenta for $\ell=0-3$. The average resonance parameters from Table~\ref{tab:averageparam} were used for $\ell=0$ and 1 contributions, in particular $S_\ell$ and $\langle \Gamma_\gamma\rangle_\ell$ were considered constant over the shown interval, while $D_\ell$ decreased exponentially as in the constant-temperature nuclear level density model, i.e. $D_\ell \propto \exp(-E_n/kT)$ with $kT=0.89$~MeV~\cite{Egidy05}. As evident from Fig.~\ref{fig:URR}, also the cross section induced by neutrons with $\ell=2$ and 3 is not negligible, especially above about 100 keV. To obtain these two components, an estimate of the average parameters for $d$- and $f$-wave resonances was necessary. The adopted $S_2=0.6\times 10^{-4}$ and $S_3=4.0\times 10^{-4}$ correspond to values for low neutron energies obtained from TALYS 2.0~\cite{TALYS} default model. Average radiation widths were taken equal to those obtained from resonances with the same parity for $\ell=0$ and 1, i.e. $\langle \Gamma_\gamma\rangle_2=104$ meV and $\langle \Gamma_\gamma\rangle_3=158$ meV.

For determination of resonance spacing we considered no parity dependence of level density. Based on $D_0$ and $D_1$ reported in Table~\ref{tab:averageparam}, we adopted $D_{J=1/2}=1470$~eV and $D_{J=3/2}=670$~eV. For higher spins we considered the spin cut-off parameter $\sigma_c=0.98A^{0.29}$ suggested for the constant temperature level density in Ref.~\cite{Egidy05}, which yields $D_{5/2}\approx 540$~eV and $D_{7/2}\approx 520$~eV.


The resulting statistical model calculations of the cross section in the unresolved resonance region yield values that are in fair agreement with the experimental capture data, and substantially lower than existing evaluations and experimental data from Musgrove \cite{Musgrove76}. The cross section reported in the ENDF/B-VIII.1 evaluation is systematically larger than the data obtained in this work. JEFF-4.0 and JENDL4 evaluations provide values of the cross section about 10-30\% lower than ENDF/B-VIII.1, but still larger than our data, at least below 200~keV.

\section{\label{sec:MACS}Astrophysical implications}
The cross section determined using the resonance parameters in Secs.~\ref{sec:RSA} and ~\ref{sec:URR} is convoluted with a Maxwellian neutron energy distribution to obtain the so-called stellar cross section or MACS, defined as
\begin{equation}
\label{macs_eq}
\textrm{MACS}=\frac{2}{\sqrt{\pi}}\frac{1}{(kT)^2}\int_{0}^{\infty} E_n\sigma(E_n)\textrm{exp}\bigg(-\frac{E_n}{kT}\bigg) dE_n.
\end{equation}
The results are listed in Table~\ref{tab:macs} for thermal energies between $kT = 5$ and 100 keV, including the specific values for the common s-process sites, e.g., for $kT = 8$ and 23 keV related to He-shell burning in low-mass AGB stars and $kT = 25$ and 90 keV for the case of core-He and shell-C burning in massive stars. 
The value for $kT = 30$ keV is used for comparison with KADoNiS~\cite{Bao2000}. In this study, the direct radiative capture component is considered to be negligible.
To evaluate the astrophysical impact of the newly determined $^{94}\text{Mo}(\mathrm{n},\gamma)^{95}\text{Mo}$ reaction rate, we performed nucleosynthesis calculations of low-mass AGB stars~\cite{Busso1999ARA&A..37..239B} using the \textsc{FuNS} evolutionary code~\cite{Cristallo2009ApJ...696..797C,Vescovi2020ApJ...897L..25V,Vescovi2021A&A...652A.100V}. Specifically, we considered stellar models with initial masses of $M = 2$ and $3\,M_{\odot}$ and a metallicity of $Z=0.01$.

In the framework of galactic chemical evolution, $^{94}\text{Mo}$ is primarily classified as a $p$-nucleus, i.e., a proton-rich isotope not produced by neutron-capture processes~\cite{Arnould2003PhR...384....1A}. However, cosmochemical evidence derived from the analysis of presolar SiC grains indicates a minor $s$-process contribution to its total abundance, estimated at approximately $4\%$~\cite{Stephan2019ApJ...877..101S}. Consistent with this picture, a net production of this isotope is predicted to occur in AGB stars~\cite{Lugaro2003ApJ...593..486L}.

Despite the significant reduction in the new MACS---approximately 25\% lower than previous evaluations~\cite{Bao2000}---the used models reveal only modest variations in the final surface abundance of $^{94}\text{Mo}$, with an increase of $3$--$4\%$ with respect to the yield obtained with the reference neutron capture cross section.
This limited sensitivity arises from the dynamics at the $^{94}\text{Nb}$ branching point ($t_{1/2} = 2.04 \times 10^4$ yr)~\cite{Szanyi2025A&A...697A..48S}. During the interpulse phase, the $^{13}\text{C}(\alpha,\text{n})^{16}\text{O}$ reaction releases neutrons over thousands of years at moderate densities ($n_n \sim 10^7$ cm$^{-3}$). Under these conditions, $^{94}\text{Nb}$ primarily decays to $^{94}\text{Mo}$, but the extended exposure time leads to nearly complete destruction of this $^{94}\text{Mo}$ via neutron captures, independently of the neutron-capture cross-section value. During subsequent thermal pulses, the $^{22}\text{Ne}(\alpha,\text{n})^{25}\text{Mg}$ source drives neutron densities to much higher values ($n_n \gtrsim 10^{10}$ cm$^{-3}$), pushing the $^{94}\text{Nb}$ branching toward neutron capture rather than decay and effectively shutting off $^{94}\text{Mo}$ production. 
Since the final abundance is bottlenecked by limited production at the $^{94}\text{Nb}$ branching point rather than by destruction efficiency, even a substantial reduction in the $^{94}\text{Mo}(\text{n},\gamma)$ cross section translates into only a marginal increase of the final yield.

A comprehensive comparison with the isotopic patterns observed in presolar SiC grains~\cite{Stephan2019ApJ...877..101S,Liu19} requires a complete set of neutron-capture cross sections for all the molybdenum isotopes and related branching-point nuclei. Recent measurements of the $^{94}$Nb$(\mathrm{n},\gamma)^{95}$Nb reaction at n\_TOF~\cite{Balibrea}, combined with the present $^{94}$Mo$(\mathrm{n},\gamma)$ results, significantly reduce the nuclear physics uncertainties in this critical branching region. In this context, the upcoming results from the ongoing analysis of $^{96}$Mo$(\mathrm{n},\gamma)$ data at n\_TOF will be crucial to complete the picture of molybdenum neutron-capture rates, refine the predicted isotopic ratios, and provide a more robust test of the $s$-process contribution to the solar system molybdenum inventory. With these experimental constraints in place, the temperature-dependent $^{94}$Nb $\beta$-decay rate under stellar conditions will remain the primary source of uncertainty for a complete understanding of the $^{94}$Mo production in AGB stars.

\begin{table}[h] 
\caption{Maxwellian-averaged capture cross sections of $^{94}$Mo (in mb) for different temperatures compared with the values in the KADoNiS database~\cite{Bao2000}.}
\begin{tabular*}{\columnwidth}{@{\extracolsep{\fill}}ccccc}
\hline \label{tab:macs}
Thermal & \multicolumn{3}{c}{This work} & KADoNiS\\
energy & Resonances & URR & Total&\\
(keV) & $E_n < 76$ keV & $76-250$ keV & &\\
\hline
5& $237(9)$&--&$237(9)$&321\\
8& $180(7)$&--&$180(7)$&\\
10& $156(6)$&$<0.3$&$156(6)$&200\\
15& $118(4)$&$2.0(1)$&$120(4)$&154\\
20& $95(4)$&$6.0(3)$&$101(4)$&128\\
23& $84(3)$&$8.0(4)$&$92(3)$&\\
25&$77(3)$&$10.0(5)$&$87(3)$&113\\
30&$65(2)$&$13(1)$&$78(3)$&$102(20)$\\
40&$48(2)$&$19(1)$&$67(2)$&89\\
50&$36(1)$&$24(1)$&$60(2)$&81\\
60&$28(1)$&$27(1)$&$55(2)$&77\\
70&$23(1)$&$29(2)$&$52(2)$&\\
80&$19(1)$&$30(2)$&$49(2)$&71\\
90&$15.4(6)$&$31(2)$&$46(2)$&\\
100&$13.1(5)$&$31(2)$&$44(2)$&68\\
\hline
\end{tabular*}
\end{table}

\section{\label{sec:conclusion}Conclusions}




The $^{94}$Mo(n,$\gamma$)$^{95}$Mo cross section has been measured with high resolution at n\_TOF and consistently combined with transmission data from GELINA within an $R$-matrix analysis up to $E_n = 32$~keV. A new and self-consistent set of neutron resonance parameters has been extracted, including improved determinations of neutron and radiative widths as well as their correlations. A total of 186 resonances were observed in this analysis of which 127 reported here for the first time. These results lead to revised average resonance properties that differ significantly from those adopted in current evaluated libraries.

At higher energies, the cross section in the unresolved resonance region, obtained from the resonance parameters and by numerical integration, was compared with Hauser--Feshbach calculations based on the extracted average parameters (i.e. $D_0=1470(260)$~eV, $D_1=460(80)$~eV, $\langle{\Gamma}_\gamma\rangle_0=104(4)$~meV, $\langle{\Gamma}_\gamma\rangle_1=158(7)$~meV, $S_0=0.48(15)\times10^{-4}$ and $S_1=4.8(9)\times10^{-4}$) and the average quantities estimated for $\ell=2$ and $\ell=3$. The extracted cross section is found to be systematically lower than existing evaluations and previous experimental measurements.


As a consequence, the resulting Maxwellian-averaged cross sections are consistently lower than previous recommendations over the entire temperature range. At $kT = 30$~keV, a value of $\mathrm{MACS} = 78 \pm 3$~mb is obtained, corresponding to a reduction of approximately 25\% with respect to earlier compilations. Similar reductions are observed at temperatures relevant to the s-process in AGB stars.

Despite this substantial decrease in the neutron-capture rate, stellar nucleosynthesis calculations indicate that this new neutron-capture rate produces only modest changes ($\sim 3$--$4\%$) in  the final $^{94}$Mo abundances. The isotopic budget is therefore primarily  controlled by the branching at $^{94}$Nb rather than by neutron capture on  $^{94}$Mo. The present results significantly reduce the nuclear-physics  uncertainties in this key region and provide improved constraints for  modeling molybdenum isotopic patterns in stellar and cosmochemical studies.



\begin{acknowledgments}
This project has received funding from the ARIEL project ("Accelerator and Research reactor Infrastructures for Education and Learning") of the European Union’s Horizon Europe Research and Innovation programme under Grant Agreement No 847594; as well as from EURO-LABS (European Laboratories for Accelerator Based Sciences) of the European Union’s Horizon Europe Research and Innovation programme under Grant Agreement No 101057511. Part of this work was supported by AEI/10.13039/501100011033 under grants PID2022-138297NB-C21 and Severo Ochoa CEX2023-001292-S. This work was supported by the Italian Ministry of University and Research (MUR) under the PRIN 2022 program, project titled "RADAMES" (grant number: 2022BAYM54). The authors acknowledge support from all the funding agencies of participating institutions.
\end{acknowledgments}

\bibliography{apssamp}

\end{document}